\documentclass[11pt]{article}
\usepackage[a4paper, left=20mm, right=20mm, top=20mm, bottom=20mm]{geometry} 
\usepackage[symbol]{footmisc}
\usepackage{tabularx}
\usepackage{booktabs}
\usepackage{amsmath, mathtools} 
\usepackage{changes} 
\usepackage{cancel} 
\setdeletedmarkup{\cancel{#1}} 
\usepackage{graphicx} 
\usepackage{setspace} 
\usepackage{xcolor} 
\newcommand{\revisioncolor}{\textcolor{black}}
\usepackage[T1]{fontenc} 
\usepackage{caladea} 
\usepackage{fancyhdr} 
\usepackage[colorlinks=true, urlcolor=blue, linkcolor=blue, citecolor=blue, bookmarksnumbered=true, bookmarksopen=true]{hyperref} 
\usepackage{authblk} 
\usepackage{cite} 
\usepackage[labelfont=bf, labelsep=period]{caption}
\usepackage{subcaption}
\usepackage{lineno} 
\usepackage{titlesec} 
\titleformat{\section}{\normalsize\bfseries}{\thesection}{1em}{} 
\titleformat{\subsection}{\normalsize\bfseries}{\thesubsection}{1em}{} 
\titleformat{\subsubsection}{\normalsize\bfseries}{\thesubsubsection}{1em}{} 
\usepackage{amssymb}

\makeatletter
\newcommand{\undersim}[1]{\mathrel{\mathpalette\@undersim{#1}}}
\newcommand{\@undersim}[2]{%
  \vcenter{%
    \ialign{%
      ##\cr
      $\m@th#1#2$\cr
      \noalign{\nointerlineskip\kern.2ex}
      $\m@th#1\sim$\cr
      \noalign{\kern-.4ex}
    }%
  }%
}
\newcommand{\gsim}{\undersim{>}}
\newcommand{\lsim}{\undersim{<}}
\makeatother

\title{\large \bfseries{Quantitative Magnetohydrodynamic Modelling of Flux Pumping in ASDEX Upgrade}}
\author[1, $\ast$]{Haowei Zhang}
\author[1]{Matthias Hölzl}
\author[2,3]{Isabel Krebs}
\author[1]{Andreas Burckhart}
\author[1]{Alexander Bock}
\author[1]{\\Sibylle Günter}
\author[1]{Valentin Igochine}
\author[1]{Karl Lackner}
\author[1]{Rohan Ramasamy}
\author[1]{Hartmut Zohm}
\author[4]{\\the JOREK team}
\author[5]{the ASDEX Upgrade team}
\affil[1]{Max-Planck-Institut für Plasmaphysik, Garching, Germany}
\affil[2]{Science and Technology of Nuclear Fusion, Department of Applied Physics, Eindhoven University of Technology, Eindhoven, The Netherlands}
\affil[3]{Dutch Institute for Fundamental Energy Research, Eindhoven, The Netherlands}
\affil[4]{See author list of Ref. \cite{Matthias2024NF}}
\affil[5]{See author list of Ref. \cite{Zohm2024AUG}}
\affil[$\ast$]{E-mail: \href{mailto:haowei.zhang@ipp.mpg.de}{haowei.zhang@ipp.mpg.de}}
\date{} 

\begin{document}


\maketitle
\noindent\textbf{Abstract: }

The sawtooth-free hybrid scenario has been achieved recently in ASDEX Upgrade (AUG) with applied non-inductive current sources and auxiliary heating [\textit{A. Burckhart et al 2023 Nucl. Fusion 63 126056}]. Control experiments in AUG suggest that the self-regulating magnetic flux pumping mechanism, characterized by anomalous current redistribution, is responsible for clamping the central safety factor ($q_0$) close to unity, thereby preventing the sawtooth onset. This work presents a numerical and theoretical investigation of flux pumping in the AUG hybrid scenario based on the two-temperature, visco-resistive, full magnetohydrodynamic (MHD) model with the JOREK code. To quantitatively model the flux pumping, we choose realistic parameters, plasma configurations, and source terms based on AUG experiments. During the initial saturation stage of the unstable 1/1 quasi-interchange mode (on millisecond timescales, \revisioncolor{i.e., thousands of Alfvén times}), $q_0$ exhibits fast under-damped oscillation and reaches a value closer to unity, which is attributed to the self-regulation of core plasma and the fast dynamo effect on the order of V/m. On the longer resistive diffusion timescale of seconds \revisioncolor{(millions of Alfvén times)}, the slow negative dynamo effect on the order of mV/m induced by the 1/1 MHD instability plays an effective role in flux pumping, which provides quantitative agreement with experimental observations for the first time. The final saturated 1/1 MHD instability exhibits features of the quasi-interchange mode and tearing mode, and the associated convective plasma flow velocity is a few m/s. The toroidal negative electric field from the slow dynamo dominantly offsets the positive current drive and continuously redistributes the current density and pressure. As a result, $q_0$ is maintained close to unity due to the low-shear profiles of current density and pressure in the plasma core, and the system enters into a sawtooth-free and quasi-stationary helical state.

\vspace{0.2cm}


\newpage 

\section{Research background}
A high design priority for the next generation of tokamak fusion reactors, such as ITER and DEMO, is sawtooth control for maintaining long-pulse discharges \cite{Chapman2013NFITER,Siccinio2022DEMO}. In the conventional inductive H-mode scenario with a radially monotonic safety factor ($q$) profile, sawtooth oscillations are normally difficult to avoid when the plasma current density and pressure are peaked over critical values to trigger the m/n = 1/1 internal kink instabilities (m and n represent the poloidal and toroidal mode number, respectively) \cite{Goeler1974PRL,Kadomtsev1975sawtooth}. These peaking behaviours result in the central safety factor ($q_0$) dropping below unity and the entrance of a $q = 1$ rational surface. Then, the tearing instability is driven by the internal kink mode and generates a 1/1 magnetic island through magnetic reconnection process. Based on Kadomtsev’s picture\cite{Kadomtsev1975sawtooth}, when the magnetic flux inside the $q=1$ rational surface is fully reconnected, the 1/1 magnetic island replaces the original core to recover the axisymmetric state, where $q_0$ is lifted close to unity, and the plasma pressure is redistributed by a fast core collapse. In Ohmic discharges or in the presence of auxiliary heating and non-inductive current drive, a new cycle of sawtooth begins with a slow rise of core current density and pressure, thereby allowing $q_0$ to drop below unity again. Despite the limitations of the Kadomtsev model (single-fluid, slow reconnection rate based on the Sweet-Park model) in explaining the fast crash timescale (tens of microseconds) and some other experimental observations\cite{Igochine2023PoP}, the majority of theoretical and numerical studies of giant sawteeth agree on the necessity of a $q = 1$ rational surface and address the fast reconnection process\cite{Wesson1990NFinertia, Park1990NFneoEta, Halpern2011PoPSawtooth, Yu2015NFsawtooth, Zhang2020Hall}. Uncontrolled sawtooth crashes can trigger neoclassical tearing modes (NTMs)\cite{Shi2015PSTntm}, degrade energy confinement, and even result in major disruptions, thus posing great challenges to plasma control in future reactor-scale tokamaks.

Besides the conventional sawtoothing discharges described above, the sawtooth-free hybrid scenario \cite{Sips2005PPCFITER, Jardin2015PRL, Krebs2017Thesis}, characterized by broad and shear-free distributions of $q$ and current density in the plasma core, has been achieved in different tokamaks \cite{Burckhart2023NF, Chapman2010NF, Piovesan2017NF, Boyes2024NF, Burckhart2024EPS} with a considerable share of non-inductive current and improved plasma confinement. Compared with the fully non-inductive reversed shear scenario, the hybrid scenario is more robust as it does not require fine control of the current profile. In contrast, the broad current density in the hybrid scenario is maintained by the magnetic flux pumping mechanism in the presence of core magnetohydrodynamic (MHD) activity \cite{Petty2009PRLfluxpumping}. Flux pumping in tokamaks is a self-regulating process of toroidal magnetized plasma under external drives, where the anomalous redistribution of current density and pressure sustains a shear-free $q$-profile ($q_0\simeq1$) in the plasma core and relaxes the plasma into a more stable helical state to prevent sawtooth onset \cite{Burckhart2023NF}. Theoretically, the current redistribution originates from the dynamo effect, which is thought to be generated by the cross-field motion of conducting plasma (MHD dynamo). The dynamo theory can be characterized by origins, including the MHD dynamo,  diamagnetic dynamo, Hall dynamo, etc.; and by timescales of action (compared with the resistive diffusion time), including the fast dynamo and slow dynamo \cite{Rincon2019JPPdynamotheory,JI2001alphaDynamoRFP}. It has been extensively studied to explain the self-generation of magnetic fields in astrophysical systems \cite{Rincon2019JPPdynamotheory} and the self-regulation process of laboratory plasmas, such as the helical deformation of the magnetized plasma in reversed-field pinch (RFP) \cite{JI2001alphaDynamoRFP, Bonfiglio2005prlRFP, Cappello2006PoPRFP, Cappello2011NFRFP, King2011PoPRFP, Piovesan2017NF}. 
For flux pumping discharges in tokamaks, the MHD dynamo effect has been extended to qualitatively explain the redistribution of plasma current density and magnetic flux in the plasma core by MHD instabilities, which prevents $q_0$ from dropping substantially below unity \cite{Jardin2015PRL}.

Specifically, linear theoretical analysis for equilibria with shear-free $q$ profiles ($q_0\simeq1$) predicts the destabilization of the 1/1 quasi-interchange mode driven by the pressure gradient\cite{Wesson1986PPCF}, which produces a convection cell in the plasma core and flattens the pressure profile. On this basis, nonlinear full MHD simulations show that the negative dynamo effect, generated by the correlated n = 1 fluctuations of plasma flows and magnetic fields, is found to be responsible for the redistribution of current density and clamps $q_0$ around unity\cite{Jardin2015PRL}. The following study predicts a threshold of the plasma beta ($\beta$) that depends on the plasma current, above which a sufficiently strong negative dynamo loop voltage (or equivalently, toroidal dynamo electric field) arises and ensures efficient flux pumping \cite{Krebs2017PoPFluxPumping}. The dependencies of flux pumping on plasma current and $\beta$ were further verified in recent ASDEX Upgrade (AUG) experiments \cite{Burckhart2023NF}, in which different phases with and without sawteeth were successfully reproduced in the same discharge by sequentially adjusting NBI (Neutral Beam Injection) power and co-ECCD (Electron Cyclotron Current Drive) intensity. Consequently, AUG flux pumping experiments provide an ideal validation reference for theoretical and numerical modelling studies. 

As discussed above, sawteeth are critical triggering mechanisms for NTMs that can cause disruptions, and they are also expected to be problematic in DEMO-like devices, as the auxiliary control of the plasma current profile is limited. In contrast, the sawtooth-free hybrid scenario with flux pumping is of great interest due to its intrinsic advantages in improved plasma confinement [allowing $\beta_N\sim3$, where $\beta_N\left(\equiv\beta a B_T/I_p\right)$ is the normalized plasma beta, $a$ the minor radius, $B_T$ the toroidal magnetic field, $I_p$ the total plasma current] and better robustness due to the nature of self-regulation, as well as potentially higher efficiencies of non-inductive current drive \cite{Burckhart2023NF}.
Similar to the AUG experiments, flux pumping has been confirmed in DIII-D discharges in the presence of a helical core induced by the external n = 1 field and a co-existing 3/2 tearing mode but without transient events like edge-localized-modes (ELMs) \cite{Piovesan2017NF} and recently even in negative triangularity (NT) and ITER baseline scenario (IBS) plasmas \cite{Boyes2024NF}. Furthermore, sawtooth-free discharges accompanied by the continuous n = 1 mode have been achieved in MAST\cite{Chapman2010NF} and EAST\cite{Mao2023PRR}, where the former mainly exhibits a low reversed shear $q$ profile with $q_\text{min}>1$, but the latter suggests the presence of the $q=1$ rational surface and a 1/1 internal kink-like mode. Recently, JET experiments presented promising progress in reproducing flux pumping in the larger device, where both the central 1/1 mode and sporadic off-axis activity were observed, but without clear sawtooth crashes \cite{Burckhart2024EPS}. These experimental results regarding the hybrid scenario and sawtooth suppression, along with subsequent analyses, demonstrate the universality of the flux pumping mechanism in tokamak plasmas. Nonetheless, the operating conditions and MHD activities to achieve the sawtooth-free hybrid scenario vary in different devices, which naturally makes it difficult to extend the parameter regimes of flux pumping in existing tokamaks and to predict future ones.

At present, the understanding of fundamental properties of flux pumping has been significantly advanced by various efforts, such as the reconstruction of the fluctuation-induced electrostatic potential and dynamo loop voltage based on experimental diagnostics \cite{Piovesan2017NF, Mao2023PRR}, the qualitative 3D MHD simulations mentioned above and some others \cite{Halpern2011PoPSawtooth, Zhang2020NFfluxpumping, Piovesan2017NF, Yu2024NFfp}. However, due to the high demands for both the computational resources and the numerical stability of MHD simulations on the resistive diffusion timescale, 3D quantitative MHD modelling of flux pumping with realistic parameters (resistivity, viscosity, heat diffusion anisotropy, etc.) for existing experimental results remains a significant challenge but is essential before we can assess the feasibility of flux pumping in future large tokamaks.

As a first quantitative milestone, this work investigates the flux pumping experiments in AUG using the JOREK code\cite{Matthias2024NF, Pamela2020PoPfMHD}. To accurately model the nonlinear plasma dynamics including the MHD dynamo effect in the AUG flux pumping discharge \cite{Burckhart2023NF}, we adopt reconstructed experimental profiles and realistic parameters. Based on the two-temperature, visco-resistive, full MHD model, simulations have been run over a resistive diffusion timescale of seconds. Two self-regulating stages of core plasma are observed, respectively caused by the fast and slow dynamo effects \cite{JI2001alphaDynamoRFP}. Firstly, at the initial saturation stage of the unstable 1/1 quasi-interchange mode (on millisecond timescales, \revisioncolor{i.e., thousands of Alfvén times}), $q_0$ shows rapid and under-damped oscillation at 5.8 kHz, and the system self-regulates into a state where $q_0$ is closer to unity. The oscillation of $q_0$ is mainly caused by the fast dynamo effect on the order of V/m, which acts much faster than resistive diffusion and is mainly generated by the quasi-interchange. Secondly, further simulations on the resistive diffusion timescale of seconds \revisioncolor{(millions of Alfvén times)} verify the effective role of the slow negative dynamo effect in flux pumping, where the associated 1/1 mode exhibits the features of both ideal quasi-interchange and resistive tearing, and the convective plasma flow velocity is a few m/s. The amplitude of the obtained negative toroidal dynamo electric field ($\sim$ mV/m) is comparable in magnitude to the experimental observations, while previous qualitative results show substantial overestimation, primarily attributed to the choice of increased resistivity \cite{Krebs2017PoPFluxPumping, Burckhart2023NF}. Especially, the slow negative dynamo can self-regulate its profile and amplitude within a certain range to counteract the positive current drive and ensure the continuous redistribution of current density and pressure in the plasma core, thereby keeping $q_0$ close to unity and leading to a sawtooth-free quasi-stationary helical state.

The remainder of this paper is organized as follows. In Sec. \ref{secSetup}, we introduce the initial simulation conditions including a brief review of the AUG flux pumping experiment, the MHD model used in JOREK, the reconstructed equilibrium and profiles, and the simulation parameters. Sec. \ref{secResults} presents the main nonlinear simulation results for demonstrating flux pumping in AUG, including the self-regulation of core plasma induced by the fast dynamo effect during the saturation stage of quasi-interchange mode and detailed comparisons of 2D (n = 0, axisymmetric and without MHD instabilities) and 3D (non-axisymmetric and with n > 0 MHD instabilities) simulation results. Sec. \ref{secDynamo} examines the role of the slow dynamo effect in the flux pumping process. The redistribution mechanisms of plasma current density and pressure by the slow dynamo are identified, which prevent the occurrence of sawteeth. Sec. \ref{secScan} presents the preliminary parameter scan of the current source, which aims to study the impact of its intensity and profile on flux pumping. In the end, the findings of the present work and the outlook for the future work are summarized and discussed in Sec. \ref{secSummary}.

\section{Experimental result and simulation setup} \label{secSetup}
\subsection{A brief review of the AUG experiment}
\begin{figure}[htbp]
	\centering
    \includegraphics[height=0.35\textwidth]{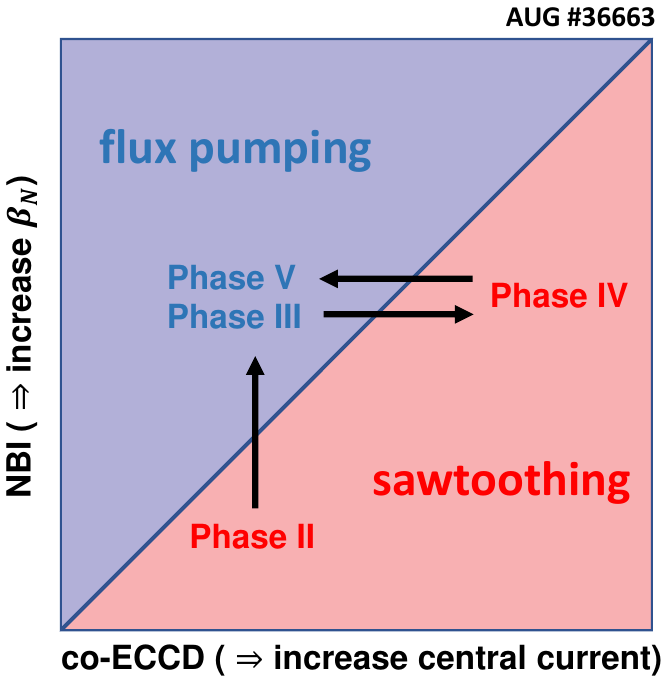}
	\caption{Schematic of the AUG discharge \#36663 \cite{Burckhart2023NF}. $\beta_N$ and plasma current are modified by adjusting NBI and co-ECCD, respectively, to reproduce the flux pumping (III and V) and sawtoothing (II and IV) phases in the same discharge. Phase I is the initial current ramp-up stage and is not plotted here.}
	\label{fig36663}
\end{figure}

JOREK \cite{Huysmans2007NFjorek, Hoelzl2021NF} simulations are carried out to investigate the flux pumping phases observed in the AUG discharge \#36663 \cite{Burckhart2023NF}. This discharge is mainly designed to experimentally study the transition behaviour between sawtoothing and flux pumping phases by adjusting the intensities of NBI heating and central co-ECCD, as shown by the schematic in Fig. \ref{fig36663}. The detailed time traces of this discharge can be found in Fig. 3 of Ref. \cite{Burckhart2023NF}. \revisioncolor{Note that the horizontal and vertical axes of the schematic only represent the relative changes in the strength of co-ECCD and NBI, respectively, and do not represent any actual values. The schematic also does not indicate any plasma behaviour in the absence of co-ECCD and NBI.}

In this experiment, the plasma transition starts from phase II with sawteeth, in which $\beta_N$ is about 1.9 and co-ECCD is less than 0.10 MA. On this basis, $\beta_N$ is increased to 2.9 by a NBI ramp-up in phase III and then remains almost constant until the end of the discharge. During phase III, sawtooth suppression is achieved by the identified flux pumping mechanism, i.e., the anomalous redistribution of central current density, which maintains $q_0$ close to unity. Subsequently, from phase III to phase IV, the intensity of co-ECCD is increased above 0.15 MA while $\beta_N$ remains unaltered. In phase IV, sporadic sawteeth reappear, indicating the failure of flux pumping. Finally, from phase IV to phase V, upon the co-ECCD intensity being reduced below 0.10 MA, the sawtooth-free phase with flux pumping is attained again.

The repeated transitions between sawtoothing and flux pumping phases clearly confirm previous theoretical prediction on the threshold of $\beta$ to suppress sawteeth through flux pumping, which is found to result from the dynamo effect by the pressure gradient driven 1/1 quasi-interchange mode\cite{Krebs2017PoPFluxPumping}.  However, the nature of the 1/1 mode and its role in the current redistribution observed during the flux pumping phases of the AUG discharge remain unclear \cite{Burckhart2023NF}. The JOREK simulations of flux pumping for phase III of AUG discharge \#36663 presented in this work aim to provide an intuitive understanding of the underlying dynamo effect and plasma self-regulation.

\subsection{Two-temperature full MHD model}
In previous flux pumping simulations, two dominant mechanisms responsible for the current redistribution are identified \cite{Krebs2017PoPFluxPumping}. The first is the negative dynamo associated with the convection flow and magnetic field distortion of a 1/1 quasi-interchange mode, which counteracts the peaking tendency of current density in a driven system. The second mechanism is the flattening and lifting of resistivity caused by the flattened electron temperature profile, which strengthens the resistive current diffusion in the plasma core. Since the initial electron temperature profile and its subsequent evolution in the presence of heating sources could deviate from ions, separated temperature equations are adopted for electrons and ions in JOREK simulations. \revisioncolor{Nevertheless, the Ohmic heating term (much less than the auxiliary heating power) and the equipartition terms between the temperatures of two species (causing numerical instabilities) have not been considered in this work}. Meanwhile, in simulations of the quasi-interchange mode driven by pressure gradient, parallel magnetic field perturbation is usually comparable with the normalized pressure perturbation ($B\delta B_\parallel\sim-\delta p$). Therefore, the full MHD model is required to describe the self-consistent evolution of $\delta B_\parallel$ and obtain the correct linear properties of the instability \cite{Pamela2020PoPfMHD}. \revisioncolor{The Hall term and the diamagnetic terms are not considered in the present simulations due to the significant challenge of long timescale simulations \cite{Pamela2020PoPfMHD}, which will be overcome in future work on the development of the extended MHD model.} The following normalized, two-temperature, single-fluid, visco-resistive, full MHD model is adopted in the base cylindrical coordinate system $\left(R,Z,\varphi\right)$
\begin{equation}\label{eq1induction}
    \begin{split}
        \dfrac{\partial\mathbf{A}}{\partial t}=\mathbf{v}\times\mathbf{B}-\eta\left({J}_\varphi-{S}_j\right)\hat{e}_\varphi-\nabla\Phi,
    \end{split}
\end{equation}
\begin{equation}\label{eq2momentum}
    \begin{split}
        \rho\dfrac{\partial\mathbf{v}}{\partial t}=-\rho\mathbf{v}\cdot\nabla\mathbf{v}+\mathbf{J}\times\mathbf{B}-\nabla p+\nabla\cdot\left(\nu\nabla\mathbf{v}\right)-S_\rho\mathbf{v},
    \end{split}
\end{equation}
\begin{equation}\label{eq3rho}
    \begin{split}
        \dfrac{\partial\rho}{\partial t}=-\nabla\cdot\left(\rho\mathbf{v}\right)+\nabla\cdot\left(D_\perp\nabla_\perp\rho+D_\parallel\nabla_\parallel\rho\right)+S_\rho,
    \end{split}
\end{equation}
\begin{equation}\label{eq4p}
    \begin{split}
        \dfrac{\partial p_{i(e)}}{\partial t} = -\mathbf{v}\cdot\nabla p_{i(e)} - \gamma p_{i(e)}\nabla\cdot\mathbf{v}+\nabla\cdot\left[\kappa_{\perp,i(e)}\nabla_\perp T_{i(e)}+\kappa_{\parallel,i(e)}\nabla_\parallel T_{i(e)}\right]+S_{T_{i(e)}},
    \end{split}
\end{equation}
where the magnetic field $\mathbf{B}=F\nabla\varphi+\nabla\times\mathbf{A}$, the plasma current density $\mathbf{J}=\nabla\times\mathbf{B}$, and the total pressure $p=\rho\left(T_i+T_e\right)$. $\mathbf{A}$ is the magnetic vector potential, $\Phi$ is the electrostatic potential, $\mathbf{v}$ is the plasma velocity, $\rho$ is the mass density, $p_{i(e)}$ is the ion (electron) pressure, $T_{i(e)}$ is the ion (electron) temperature, $F=F\left(\psi\right)$ is an axisymmetric equilibrium function of the poloidal magnetic flux $\psi$, defined to satisfy the Grad–Shafranov equilibrium (the equilibrium toroidal magnetic field $B_{\varphi,\text{eq}}=F/R$) and is kept constant over time, ${S}_j$ is the current source in the toroidal direction ($\varphi$), $S_\rho$ is the particle source, $S_{T_{i(e)}}$ is the ion (electron) heating source, $\gamma=5/3$ is the ratio of specific heats of a monatomic gas, $\eta$ is the resistivity (applied only in the toroidal direction \revisioncolor{to improve the numerical stability}), $\nu$ is the dynamic viscosity, $D_{\perp\left(\parallel\right)}$ is the perpendicular (parallel) particle diffusivity, $\kappa_{\perp\left(\parallel\right)}$ is the perpendicular (parallel) heat conductivity.

All variables are normalized mainly based on the vacuum magnetic permeability $\mu_0$ and the central mass density $\rho_0$. The detailed normalizations of quantities can be found in Ref. \cite{Hoelzl2021NF}. Nevertheless, in this work, to provide a direct comparison with the experimental observations, all simulation results will be presented in the International System of Units. In practice, the Weyls' guage is chosen to eliminate $\Phi$ in Eq. \ref{eq1induction}, that is, $\mathbf{A}^\prime=\mathbf{A}+\nabla\Psi$, $\Phi^\prime=\Phi-\partial_t\Psi$, and $\partial_t\Psi=\Phi$, then $\partial\mathbf{A}^\prime=\mathbf{v}\times\mathbf{B}-\eta\left({J}_\varphi-{S}_j\right)\hat{e}_\varphi$ \cite{Pamela2020PoPfMHD, Jackson2002AJPguage}. Detailed implementation and validation of the full MHD model in JOREK code can be found in Refs. \cite{Haverkort2016JCPjorek, Pamela2020PoPfMHD}.

\subsection{Initial equilibrium}
The flux pumping simulations are carried out for phase III of AUG discharge \#36663, as shown in Fig. \ref{fig36663}. The initial static equilibrium and source terms are reconstructed based on the diagnostic data at the time interval of 3.75 - 3.95s (see Fig. 3 of Ref. \cite{Burckhart2023NF}). When reconstructing the equilibrium, the IMSE (Imaging Motional Stark Effect) diagnostic data is included with a larger weight than the neoclassical current diffusion constraint employed in the IDE (integrated data analysis equilibrium) solver. Thus, the solution agrees well with the experimental IMSE angles over the neoclassical current diffusion model and can reflect $q$ value with a precision of 0.1. The reconstructed equilibrium with IMSE data shows that $q_0$ in the flux pumping phase is clamped around unity, corresponding to the 'experimental $q_0$' in Ref. \cite{Burckhart2023NF}, in contrast to the 'modeled' $q_0$ (lower than unity) that does not take into account the IMSE data and anomalous current redistribution.

The toroidal magnetic field on the axis $B_{0}$ is $-2.57$ T ($F_{0} = R_{0}B_{0}=-4.41$). The total plasma current $I_p$ enclosed by the last closed-flux surface (LCFS) is kept constant over time ($-0.806$ MA) with the toroidal loop voltage imposed on the boundary by proportional-integral-differential (PID) feedback control.

\begin{figure}[htb]
    \centering
    \includegraphics[width = \textwidth]{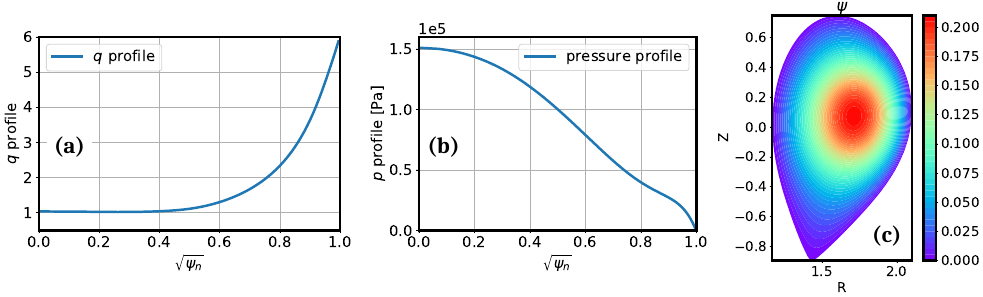}
    \caption{The radial profiles of (a) safety factor and (b) pressure, and (c) the initial distribution of poloidal magnetic flux for AUG discharge \#36663 \cite{Burckhart2023NF} at the time interval of 3.75 - 3.95s.}
    \label{figqppsi}
\end{figure}

The initial radial profiles of $q$ and pressure are plotted in Fig. \ref{figqppsi} (a) and (b). The central $q$ profile exhibits a shear-free distribution over a minor radius up to $\rho_p \approx 0.4$ in the plasma core (where $\rho_p = \sqrt{\psi_n}$, and $\psi_n$ is the normalized poloidal magnetic flux). The initial $q_0$ equals 1.04. Fig. \ref{figqppsi} (c) shows the initial distribution of $\psi$. The simulation domain is constrained within the LCFS, and the fixed boundary condition is used. With the given equilibrium, the 1/1 quasi-interchange mode driven by the pressure gradient is linearly unstable. 

\subsection{Source terms and diffusion coefficients}
To quantitatively model the flux pumping in the AUG discharge, experimental source terms and realistic diffusion coefficients, especially the resistivity and viscosity, are adopted. These considerations are necessary because the resistivity directly affects the driving strength of the current source, which is proportional to the resistivity as shown in Eq. \ref{eq1induction}. The viscosity may influence the amplitude of convection flow generated by the 1/1 mode, thereby influencing the nonlinear behaviour of the 1/1 mode \cite{Halpern2011PoPSawtooth, Zhang2020NFfluxpumping}.

\begin{figure}[htbp]
    \centering
    \includegraphics[width = 0.8\textwidth]{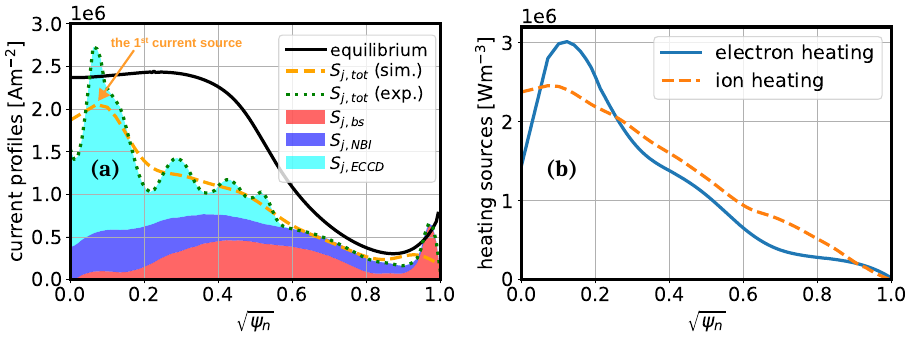}
    \caption{Radial profiles of (a) the toroidal current density of the equilibrium (solid line), the smoothed non-inductive current source used in simulations (dashed line, corresponding to the first current source in Fig. \ref{figCurrentSources}), \revisioncolor{the unmodified total experimental non-inductive current source (dotted line), and the contributions of bootstrap current (red), NBI (blue) and ECCD (cyan) to the total non-inductive current source;} and (b) heating sources for electrons (solid line) and ions (dashed line) from AUG discharge \#36663 \cite{Burckhart2023NF} at the time interval of 3.75 - 3.95s.}
    \label{figsource}
\end{figure}

The toroidal current density of the initial equilibrium and the radial source profiles of the non-inductive current drive (sum of NBI, ECCD, and bootstrap current), electron heating [sum of ECRH (Electron Cyclotron Resonance Heating) and NBI] and ion heating (NBI) are plotted in Fig. \ref{figsource}. The initial toroidal current density exhibits a flat distribution in the plasma core with a value of about 2.4 MA/m$^{2}$, which is consistent with the $q$ profile in Fig. \ref{figqppsi} (a). The original experimental current source exhibits a slightly off-axis peaked profile, maximizing the value at $\rho_p=0.1$, as shown by the dotted line in Fig. \ref{figsource} (a). The off-axis distribution of the current source mainly results from the slightly off-axis deposition of ECRH beams, which also results in a similar distribution of electron heating in Fig. \ref{figsource} (b). The ions are primarily heated by NBI, which is more smoothly distributed as shown in Fig. \ref{figsource} (b). It should be noted that for the consideration of numerical stability, the current source used in JOREK simulations is heavily smoothed here, as shown by the dashed line in Fig. \ref{figsource} (a), and the peak intensity at $\rho_p=0.1$ is smaller than that of the experimental profile (2.05 vs. 2.70 MA/m$^{2}$), but is more broadly distributed in the plasma core. The share of the Ohmic current to the total one is roughly 20\% in the plasma core. The total plasma current $I_p$ is maintained constant over time by the feedback control loop voltage applied on the simulation boundary. 
A preliminary parameter scan of the current source is presented in Sec. \ref{secScan} to study the influence of its radial profile and intensity. The particle source is not considered in the simulation ($S_\rho=0$) because the plasma density can be maintained at a relatively stationary level in the absence of perpendicular particle diffusivity in the long-term simulation.

We choose the realistic Spitzer resistivity and a semi-empirical viscosity (with the value between the Braginskii perpendicular and gyro viscosities) \cite{Vivenzi2022Viscosity}, both depending on the electron temperature, i.e., $\eta=\eta_{0}T_e^{-3/2}$ and $\nu=\nu_{0}T_e^{-3/2}$. $\eta_{0}$ and $\nu_{0}$ are the values of resistivity and viscosity on the axis. Unless otherwise specified, we use $\eta_{0} = 2.41\times 10^{-9}\ \Omega\cdot \text{m}$ for both 2D and 3D simulations. And $\nu_{0}=4.51\times 10^{-7}\ \text{kg}\cdot \text{m}^{-1}\text{s}^{-1}$ (kinematic viscosity $\Tilde{\nu}_{0}=2.70\ \text{m}^2/\text{s}$). The corresponding magnetic Prandtl number is $1400$, the Hartmann number is $7.90\times 10^6$ \footnote[2]{Note that due to the low resistivity and viscosity, the Hartmann number in this work is much larger than that in previous MHD simulations for tokamak and RFP plasmas \cite{Shen2018NF, Zhang2020NFfluxpumping, Krebs2017PoPFluxPumping, Cappello2000PRL_RFP, Bonfiglio2005prlRFP, Cappello2006PoPRFP}. As will be discussed in Subsec. \ref{subsecOutlook}, the preliminary parameter scan shows that increasing the Hartmann number results in the transition of core plasma from flux pumping towards sawtooth-like oscillation and saturated resistive internal kink mode. Detailed simulation results of the viscosity and resistivity dependence of different plasma states will be reported in a separate paper.}, and the Lundquist number is $2.80\times10^8$ (characteristic length $L=0.1\ \text{m}$) \cite{Cappello2000PRL_RFP}. Realistic temperature-dependent parallel heat conductivities are respectively implemented for ions and electrons, i.e., $\kappa_{\parallel,i(e)}=\kappa_{\parallel,i(e),0} T_{i(e)}^{5/2}$. $\kappa_{\parallel,i(e),0}$ is calculated based on the ion (electron) temperature on the axis using Spitzer-Haerm formula \cite{SpitzerHaerm1953PRkappa} (at high temperature, the parallel heat conductivities are typically overestimated, and they are reduced by a factor of 30 due to the heat ﬂux limit correction\cite{Yu2000PoP, Matthias2010Thesis}), where $\kappa_{\parallel, i, 0}=37\ \text{kg}\cdot \text{m}^{-1}\text{s}^{-1}$ and $\kappa_{\parallel, e, 0}=1174 \ \text{kg}\cdot \text{m}^{-1}\text{s}^{-1}$. The radial profiles of perpendicular heat conductivities are calculated from the axisymmetric stationary conditions for ion and electron temperatures, i.e., satisfying $\nabla\cdot\left[\kappa_{\perp,i(e)}\nabla_\perp T_{i(e)}\right]+S_{T_{i(e)}}=0$. As a result, the temperature amplitudes and profiles can be roughly maintained in the 2D simulations without MHD instabilities. The values on the axis are $\kappa_{\perp, i, 0}=2.64\times 10^{-7}\ \text{kg}\cdot \text{m}^{-1}\text{s}^{-1}$ and $\kappa_{\perp, e, 0}=1.81\times 10^{-7}\ \text{kg}\cdot \text{m}^{-1}\text{s}^{-1}$, respectively. The anisotropy of electron heat conductivity ($\kappa_\parallel/\kappa_\perp$) is of the order of $10^{10}$. In the continuity equation as Eq. \ref{eq3rho}, the perpendicular particle diffusivity $D_\perp$ is set as 0, \revisioncolor{and in order to maintain numerical stability,} a constant parallel particle diffusivity $D_\parallel=2.17\times10^5\ \text{m}^2\cdot\text{s}^{-1}$ is adopted to average density perturbations along the magnetic field line and avoid negative density values. The detailed values and distributions of the adopted parameters are also listed in Table \ref{TableSetup}.

In JOREK, the implicit Gear scheme for time advancement, the G1-continuous finite element (quadrangular bi-cubic Bezier elements) \revisioncolor{\cite{Czarny2008JCPG1, Hoelzl2021NF, PAMELA2022Gcontinous}} based on the weak form for the ($R,\ Z$) poloidal plane and a finite Fourier series expansion in the toroidal direction ($\varphi$) \cite{Haverkort2016JCPjorek} are adopted.
The time step of $2.30\times 10^{-3}\ \text{ms}$ and a polar grid with the resolution of $\left(n_\text{poloidal}, n_\text{radial}\right) = \left(100, 100\right)$ are used. Parameter scans in both time step and spatial resolution were conducted to ensure converged linear growth rates and saturation amplitudes of the 1/1 quasi-interchange. \revisioncolor{A convergence scan of different truncations on the Fourier series of toroidal modes shows that the 3D simulation with only n $\le$ 1 harmonics cannot correctly capture the flux pumping. With n $\le$ 2 and n $\le$ 4, the flux pumping can be obtained in both cases, with the relative differences in the saturated $q_0$ and toroidal current density of the order of 1\%. Therefore, the maximum toroidal mode number is set to 4 in the 3D simulations to guarantee convergence of the saturated solution under acceptable computational resources and numerical uncertainties.}

\begin{table}[thb]
  \centering
  \caption{Simulation parameters in JOREK.}
  \label{TableSetup}
  {\footnotesize{
    \begin{tabular}{p{5cm} p{5cm} p{5cm}}
      \toprule
      \bf{item} & \bf{value} & \bf{distribution} \\
      \midrule
      Spitzer resistivity & $\eta_{0} = 2.41\times 10^{-9}\ \Omega\cdot\text{m}$ \newline \revisioncolor{\ \ \ \ \ \ $ = 8.85\times 10^{-10} $ (JOREK units)} & $\eta = \eta_{0}\left(T_e/T_{e,0}\right)^{-3/2}$  \\
      Dynamical viscosity & $\nu_{0} = 4.51\times 10^{-7}\ \text{kg}\cdot \text{m}^{-1}\text{s}^{-1}$ \newline \revisioncolor{\ \ \ \ \ \ $ = 1.23\times 10^{-6} $ (JOREK units)} & $\nu = \nu_{0}\left(T_e/T_{e,0}\right)^{-3/2}$  \\
      Kinematic viscosity & $\Tilde{\nu}_0=2.70\ \text{m}^2/\text{s}$ \\
      \midrule
      Parallel heat conductivity (e) & $\kappa_{\parallel, e, 0}=1174\ \text{kg}\cdot \text{m}^{-1}\text{s}^{-1}$ \newline \revisioncolor{\ \ \ \ \ \ \ \ \ \ \ \  $ = 2136 $ (JOREK units)} & $\kappa_{\parallel, e} = \kappa_{\parallel, e, 0}\left(T_e/T_{e,0}\right)^{5/2}$  \\
      Parallel heat conductivity (i) & $\kappa_{\parallel, i, 0}=37\ \text{kg}\cdot \text{m}^{-1}\text{s}^{-1}$ \newline \revisioncolor{\ \ \ \ \ \ \ \ \ \ \ \  $ = 67 $ (JOREK units)} & $\kappa_{\parallel, i} = \kappa_{\parallel, i, 0}\left(T_e/T_{e,0}\right)^{5/2}$  \\
      Perpendicular heat conductivity (e) & $\kappa_{\perp, e, 0}=1.81\times 10^{-7}\ \text{kg}\cdot \text{m}^{-1}\text{s}^{-1}$ \newline \revisioncolor{\ \ \ \ \ \ \ \ \ \ \ \ \ $ = 3.34\times 10^{-7} $ (JOREK units)} & $\nabla\cdot\left(\kappa_{\perp, e}\nabla_\perp T_e\right)+S_{T_e}=0$  \\
      Perpendicular heat conductivity (i) & $\kappa_{\perp, i, 0}=2.64\times 10^{-7}\ \text{kg}\cdot \text{m}^{-1}\text{s}^{-1}$ \newline \revisioncolor{\ \ \ \ \ \ \ \ \ \ \ \ \ $ = 4.83\times 10^{-7} $ (JOREK units)} & $\nabla\cdot\left(\kappa_{\perp, i}\nabla_\perp T_i\right)+S_{T_i}=0$  \\
      \midrule
      Parallel particle diffusivity & $D_\parallel = 2.17\times 10^5\ \text{m}^2/\text{s}$ \newline \revisioncolor{\ \ \ \ \ \ \ \ $ = 0.1 $ (JOREK units)}& uniform \revisioncolor{(for the sake of numerical stability)} \\
      Perpendicular particle diffusivity & $D_\perp = 0$ &  \\
      \midrule
      Magnetic Prandtl number \newline  [$P\equiv\nu\mu_0/\left(\rho_m\eta\right)$] & $P_0=1400$ & \\
      Hartmann number \newline [$H\equiv\left(\mu_0L^2v_A^2/\left(\eta\Tilde{\nu}\right)\right)^{0.5}$] & $H_0=7.90\times 10^6$ ($L=0.1\ \text{m}$)  & \\
      Lundquist number \newline  ($S\equiv\mu_0Lv_A/\eta$) & $S_0=2.80\times 10^8$ ($L=0.1\ \text{m}$) & \\
      \midrule
      Central mass density ($\rho_m$) & $0.98\times 10^{20}\ \text{m}^{-3} \cdot m_\text{proton}$ & \\
      \revisioncolor{Central magnetic field} ($B_{0}$) & $-2.57$ T &  \\
      \revisioncolor{Central safety factor} ($q_{0}$) & 1.04 &  \\
      \midrule
      Target current for feedback control & $-8.06\times 10^5\ \text{A}$& \\
      Time step & $2.30\times 10^{-3}\ \text{ms}$ & \\
      \bottomrule
    \end{tabular}
  }}
\end{table}

\section{Simulation results of flux pumping} \label{secResults}
Simulations are carried out with the above initial settings both in 2D and 3D. In the 2D simulation, only the axisymmetric components (n = 0) of all variables are advanced, which is equivalent to solving the resistive current diffusion equation in the presence of the current source but in the absence of anomalous current redistribution. The temporal evolution of central current density and safety factor corresponds to the 'modeled' results in Ref. \cite{Burckhart2023NF}, where the IMSE data is not included to ignore anomalous current redistribution mechanisms besides the neoclassical current diffusion and a sawtooth current redistribution model. In this circumstance, the central peaking behaviour of toroidal current density is expected in the presence of the current source, and $q_0$ will decrease far below unity. In the 3D simulation, MHD instabilities (n > 0) are allowed to develop, and the axisymmetric components evolve self-consistently through nonlinear mode coupling. The nonlinear evolution of toroidal current density and safety factor in the plasma core will be compared between 2D and 3D simulations to identify the roles played by the MHD instability and dynamo effect in the flux pumping discharge.

\subsection{2D simulation results}
\begin{figure}[htbp]
	\centering
    \includegraphics[width=0.55\textwidth]{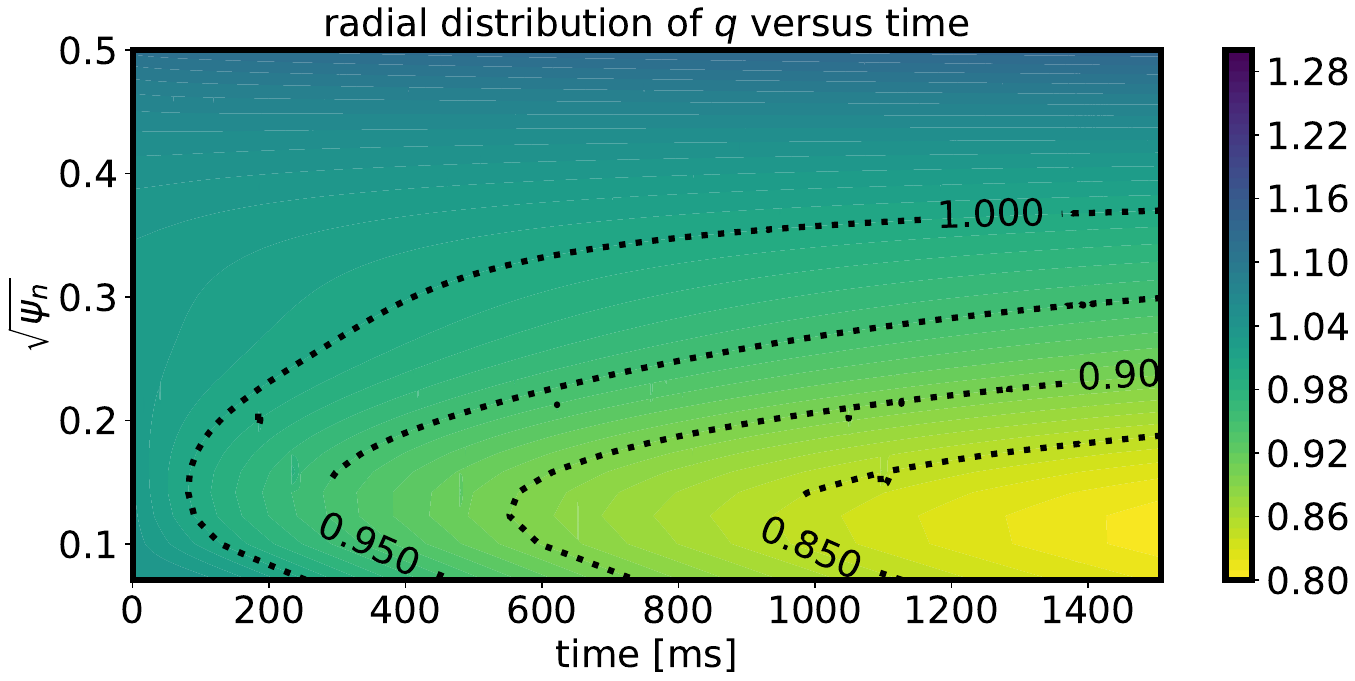}
	\caption{Temporal evolution of $q$ profile in the 2D simulation. The color map indicates the value of $q$ and the dotted lines mark the contour of several specific values.}
	\label{figq2D}
\end{figure}

\begin{figure}[htbp]
    \centering
    \includegraphics[width = 0.99\textwidth]{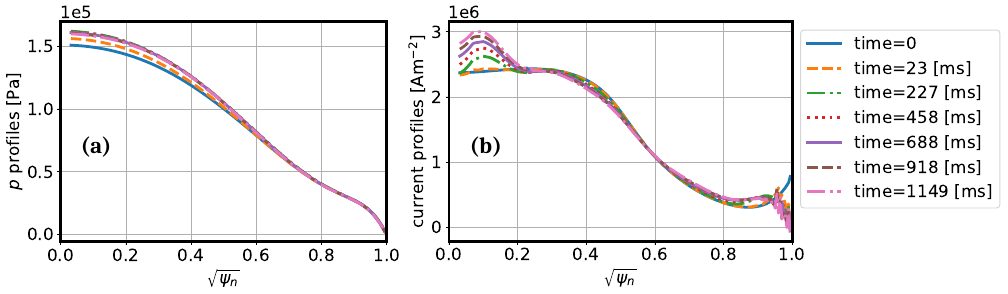}
    \caption{Radial profiles of (a) plasma pressure and (b) toroidal plasma current density at different moments in the 2D simulation.}
    \label{figpJ2D}
\end{figure}

The time trace of the $q$ profile in the 2D simulation is shown in Fig. \ref{figq2D}, where the central safety factor presents a continuous decline from 1.04 to below unity. In the simulation, we adopt relatively large perpendicular heat conductivities to balance the heating sources, as shown in Fig. \ref{figsource} (b). This way, the plasma pressure is maintained almost constant over time, as shown in Fig. \ref{figpJ2D} (a). Therefore, as shown in Fig. \ref{figpJ2D} (b), the decrease of $q_0$ is primarily caused by the peaking of current density due to the current drive instead of the heat source as in the case of Ref. \cite{Krebs2017PoPFluxPumping}. At 1500 ms, the minimum value of $q$, located at $\rho_p=0.1$, is $q_\text{min}\simeq 0.79$, and the maximum toroidal current density is increased by about 30\% compared with the initial value, i.e., $\Delta J_\varphi\approx$ 0.8 MA/m$^{2}$. The radial profiles of $q$ and $J_\varphi$ in the 2D simulation are consistent with the off-axis peaked current source as shown in Fig. \ref{figsource} (a). In Fig. \ref{figq3D} (d), we also adopt an increased resistivity ($\times 100$) to obtain the saturated $q$ profile in a much shorter computational timescale. The result indicates $q_0$ saturates around 0.75 in the 2D simulations, which is slightly higher than the value of 0.6 at 4.5 s estimated in Fig. 3 (d) of Ref. \cite{Burckhart2023NF} (see the 'modeled' curve). This difference in the saturated $q_0$ is reasonable because the current source shown in Fig. \ref{figsource} (a) has been smoothed, and as will be shown in Sec. \ref{subsecCurrentDiffusion}, the current driving strength is appropriately reduced based on the current diffusion model.

In the absence of MHD instabilities, the condition of zero plasma flow ($\mathbf{v}\simeq 0$) is well satisfied in the plasma core. Therefore, the evolutions of current density and safety factor are mainly determined by the resistivity term (proportional to $\eta$) in Eq. \ref{eq1induction}. The characteristic resistive diffusion timescale ($\tau_R=\mu_0L^2/\eta$) of core plasma current density is of the order of seconds, which is consistent with the observation regarding the decline in 'modeled' $q_0$ in Ref. \cite{Burckhart2023NF}. The 2D simulation predicts the saturated $q_0$ much lower than unity, and consequently, the sawtooth should be destabilized by the peaked current density, which contradicts the experimental observation of sawtooth-free behaviour for phase III in AUG discharge \#36663 \cite{Burckhart2023NF}. 

The failure of the 2D model in predicting evolutions of the plasma current density and safety factor suggests the importance of the 3D effect in modelling flux pumping. In the 3D simulation with non-axisymmetric MHD instabilities and axisymmetric equilibrium being self-consistently evolved, the temporal evolutions of current density and $q$ profiles are totally different from the 2D simulation, as described below.

\subsection{Linear characteristics of the MHD instability in the 3D simulation}
\begin{figure}[htbp]
    \centering
    \includegraphics[width = 0.8\textwidth]{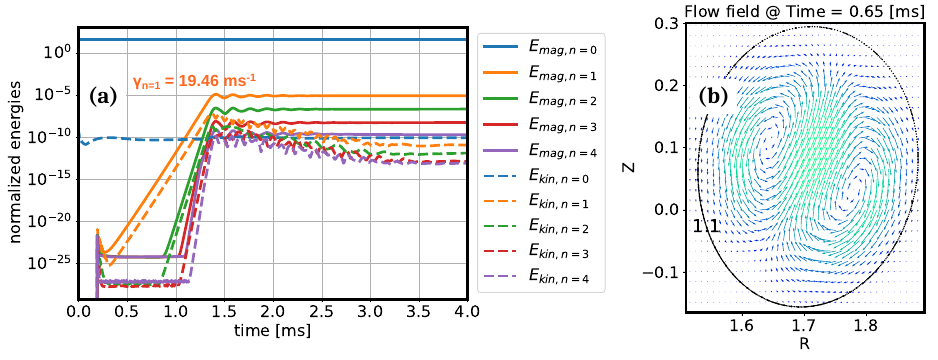}
    \caption{(a) Time trace of the normalized magnetic (solid line) and kinetic (dashed line) energies for the linear and early nonlinear stages of the 3D simulation. The linear growth rate of the n = 1 mode is 19.46 ms$^{-1}$. (b) Plasma flow in the linear stage of the n = 1 instability in the 3D simulation. The black dotted line indicates the magnetic surface with $q = 1.1$.}
    \label{fig3DEnergiesModestructureLinear}
\end{figure}

In this subsection, we first analyze the linear characteristics of the MHD instability for the equilibrium shown in Fig. \ref{figqppsi}. In the 3D simulation, toroidal harmonics are limited to n $\le$ 4, and the toroidal mode number of the most unstable instability is n = 1. The time trace of normalized magnetic and kinetic energies is plotted in Fig. \ref{fig3DEnergiesModestructureLinear} (a). During 0.25 $\sim$ 1.3 ms, the n = 1 mode grows linearly with a growth rate of 19.46 ms$^{-1}$, while the other modes with n = 2 $\sim$ 4 are not growing linearly, but are excited by the n = 1 via nonlinear mode coupling. The dominant poloidal mode number is m = 1. In Fig. \ref{fig3DEnergiesModestructureLinear} (b), the linear eigenmode structure of the n = 1 instability is plotted in the form of the poloidal flow field. Within the central low magnetic shear region, a strong convection cell forms. The plasma flow exhibits a large amplitude in the whole plasma core, which aligns with the feature of the 1/1 quasi-interchange mode and is quite distinct from that of the 1/1 internal kink mode. In the latter case, the poloidal flow dominates around the $q = 1$ rational surface. A more detailed comparison of the linear eigenmode structures of 1/1 quasi-interchange mode and 1/1 internal kink mode is presented in Fig. 14 of Ref. \cite{Krebs2017PoPFluxPumping}.

The 1/1 quasi-interchange mode is an ideal MHD instability and is driven by the pressure gradient. It should be noted that the linear growth rate of the instability is very sensitive to the initial value of $q_0$. However, the nonlinear flux pumping mechanism is a self-regulating process, and the slight deviation of the initial $q_0$ will not significantly influence the final saturated state. As shown below in the nonlinear 3D simulation, the plasma equilibrium will self-regulate to saturate at a proper helical state with a final $q_0$ close to unity to provide the toroidal dynamo electric field, which balances the driving effect from the current source. Therefore,  it can be expected that the realistic resistivity will be crucial to obtain the correct dynamo effect in the 3D nonlinear simulation, although varying resistivity has a limited impact on the linear growth rate of the instability \cite{Krebs2017PoPFluxPumping}. 

\subsection{Self-regulation and fast dynamo in the early saturation stage}

\begin{figure}[htbp]
    \centering
    \includegraphics[width = \textwidth]{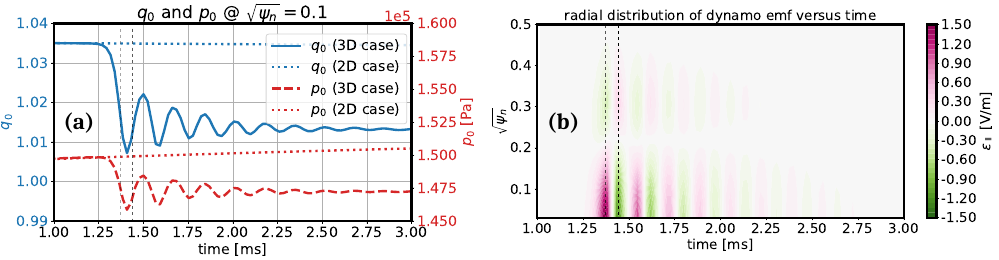}
    \caption{Temporal evolutions of (a) $q_0$ and \revisioncolor{$p_0$} in the 2D and 3D simulations, and of (b) the parallel dynamo emf profile in the 3D simulation during the early saturation stage of the 1/1 quasi-interchange mode. Two vertical dashed lines indicate the moments of the first two extreme values of the dynamo term, i.e., 1.362 ms and 1.431 ms, respectively.}
	\label{figFastdynamo3D}
\end{figure}

Before we investigate the flux pumping occurring in the resistive diffusion timescale, the rapid oscillations of $q_0$ \footnote[3]{\revisioncolor{Note that the $q$ profiles shown in this paper are mainly obtained by tracing the axisymmetric (n = 0) component of the magnetic field, with the exception of Fig. \ref{figPoincare3D}, which is calculated by tracing the total magnetic field and considering the helical magnetic axis as reference \cite{Zhang2020q_evolution}.}} \revisioncolor{and the central plasma pressure ($p_0$)} due to the fast dynamo effect in the 3D simulation are noteworthy when the 1/1 quasi-interchange mode becomes prominent in its early saturation phase, as shown in Fig. \ref{figFastdynamo3D} (a). \revisioncolor{Taking the evolution of $q_0$ as an example,} at about 1.3 ms, $q_0$ first drops rapidly by 0.03 on a short timescale of 0.1 ms. \revisioncolor{This is much faster than the evolution of $q_0$ in the 2D case that occurs in the resistive diffusion timescale.} Then $q_0$ oscillates around 1.013 with the frequency of 5.8 kHz for a few milliseconds until it becomes relatively stable. After the fast oscillation phase in Fig. \ref{figFastdynamo3D} (a), $q_0$ enters a slow decay phase with a characteristic timescale of hundreds of milliseconds, as plotted by the solid line in Fig. \ref{figq3D} (c). It will be shown in Sec. \ref{secDynamo} that the ultimate saturation of $q_0$ is mainly determined by the balance between the current peaking (due to current drive and auxiliary heating) and the negative slow dynamo effect (contributed by non-axisymmetric fluctuations of plasma velocity $\mathbf{v}_1$ and magnetic field $\mathbf{B}_1$ ), which occurs on the resistive diffusion timescale, much slower than the oscillation of $q_0$ observed here.

Fig. \ref{figFastdynamo3D} (b) shows the evolution of the parallel dynamo electromotive force (emf) in the plasma core, which is projected along the axisymmetric mean magnetic field $\mathbf{B}_0$ and averaged along the flux surface (indicated by $\langle\cdots\rangle$), i.e., $\varepsilon_\parallel=\langle\mathbf{B}_0\cdot\left(\mathbf{v}_1\times\mathbf{B}_1\right)/B_{\varphi,\text{eq}}\rangle$. The process of extracting the parallel dynamo emf from the induction equation (Eq. \ref{eq1induction}) is detailed in Sec. \ref{subSectionDynamoParallel}. During the first decreasing phase of $q_0$, the dynamo emf in the plasma core grows exponentially until 1.5 V/m. Then, it evolves to negative values and subsequently oscillates between positive and negative values with the same frequency as $q_0$. The moments of the first two dynamo extremes are labelled by two dashed lines in Fig. \ref{figFastdynamo3D}. Sec. \ref{secDynamo} will demonstrate that the positive dynamo in the core (1.36 ms, the first dashed line) mainly increases the poloidal magnetic flux, as well as the current density and the helicity of the mean magnetic field, thus reducing $q_0$. Conversely, the negative dynamo (1.43 ms, the second dashed line) pumps the poloidal magnetic flux outward from the core and raises $q_0$. The dynamo emf changes its sign around $\rho_p=0.2$, indicating the radial redistribution of magnetic flux and current density. 

The system during the initial saturation stage behaves like an under-damped pendulum \cite{benenson2006handbook}, where the oscillation amplitude of $q_0$ and the dynamo strength decrease significantly over time due to friction of the system, e.g., the viscosity. Specifically, the dynamo emf decreases from the initial V/m level to a few mV/m later. The former large amplitude dynamo ($\sim$ V/m) is referred to as the \textbf{fast dynamo}, as it changes the magnetic field and $q_0$ on a timescale much faster than the resistive diffusion \cite{JI2001alphaDynamoRFP}. As will be shown in Sec. \ref{secDynamo}, the later small amplitude dynamo ($\sim$ mV/m) redistributes magnetic flux and current density on the resistive diffusion timescale and is therefore referred to as the \textbf{slow dynamo} \cite{JI2001alphaDynamoRFP}, which plays a key role in maintaining flux pumping and the quasi-stationary state of the core plasma.

The rapidly damped oscillation of $q_0$\revisioncolor{, $p_0$}, and dynamo emf is essentially a self-regulating relaxation of the core plasma, which can potentially enhance the robustness of flux pumping discharge. In tokamak experiments, after the initial current ramp-up, the discharge will likely begin with a slight deviation of $q_0$ from the target value for maintaining the quasi-stationary flux pumping state. However, the slight error in equilibrium will be roughly reduced by the fast dynamo effect on the short saturation timescale of 1/1 quasi-interchange. Subsequently, the equilibrium will be further refined by the competition between current-driven mechanisms and the slow dynamo effect on the resistive diffusion timescale. In such a case, to avoid the sawtooth onset via flux pumping in experiments, it is not necessary to manually fine-tune the central current and pressure profiles, but the proper equilibrium with $q_0$ close to unity can be achieved by the self-regulation of the core plasma.

\subsection{Nonlinear 3D simulation until quasi-stationary state}
\begin{figure}[htbp]
	\centering
    \includegraphics[width=0.55\textwidth]{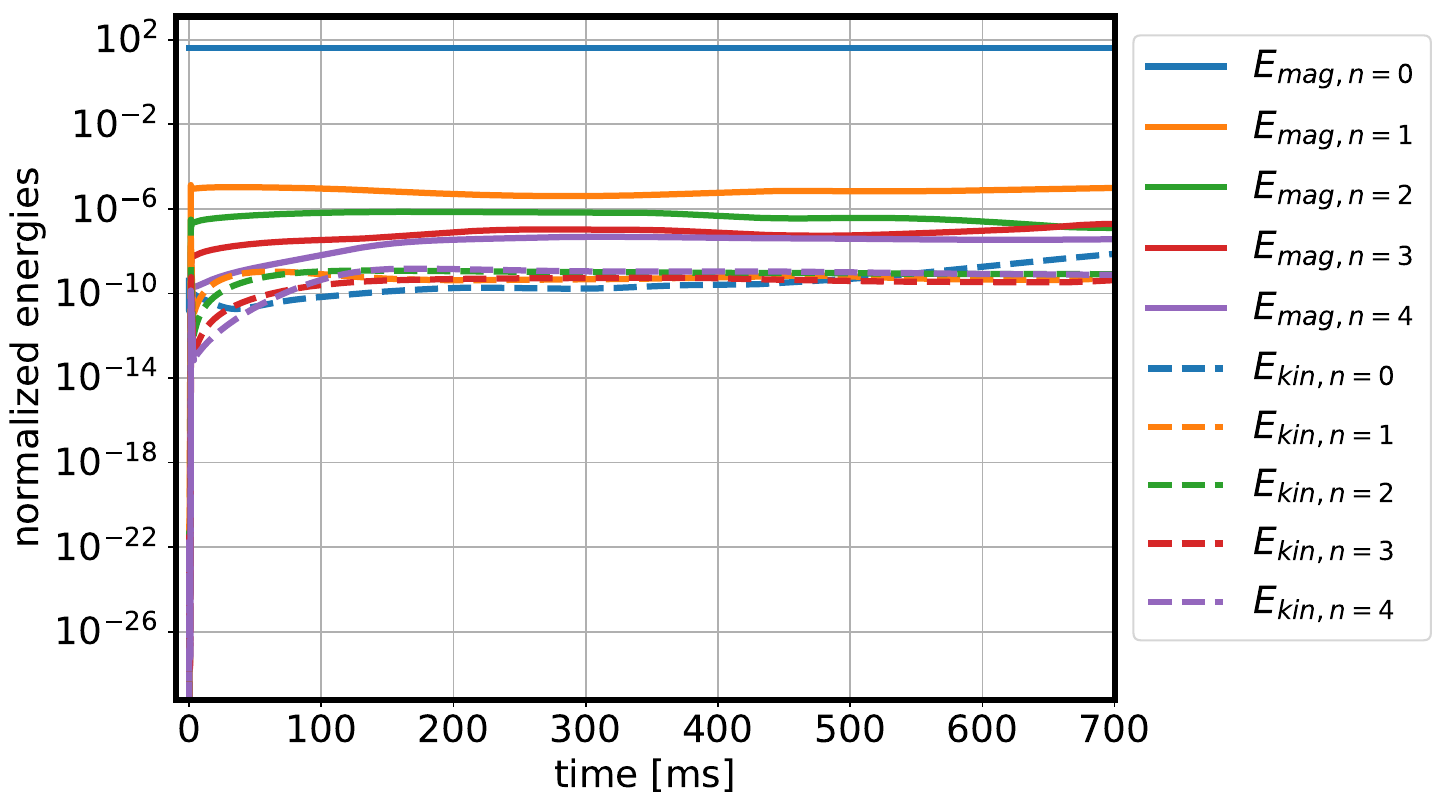}
	\caption{Time trace of the normalized magnetic (solid line) and kinetic (dashed line) energies from the nonlinear 3D simulation.}
	\label{fig3DEnergiesNonlinear}
\end{figure}

The 3D nonlinear simulation is further carried out for a considerably longer timescale of resistive diffusion to achieve the quasi-stationary state of core plasma and observe the identified phenomenon of flux pumping. The time trace of energies \revisioncolor{over 700 ms} is plotted in Fig. \ref{fig3DEnergiesNonlinear}. The system evolves in a rather stationary way, and the dominant mode throughout the simulation is the n = 1 mode. 

\begin{figure}[htb]
    \centering
    \includegraphics[width = \textwidth]{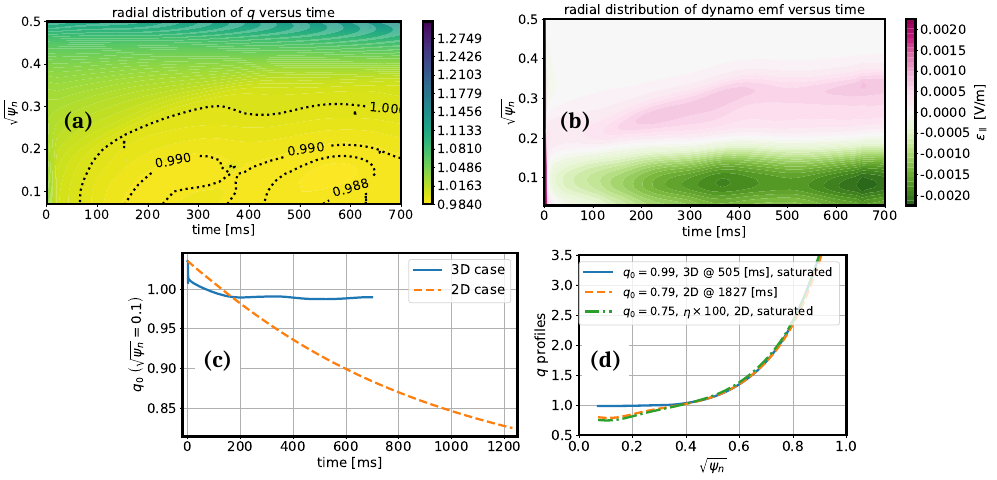}
    \caption{(a) Temporal evolution of $q$ profile in the 3D simulation, which is calculated with the axisymmetric component of the magnetic field, also for (c) and (d). The color map indicates the value of $q$ and the dotted lines mark the contour of several specific values. (b) Temporal evolution of the parallel dynamo emf profile in the 3D simulation. (c) Temporal evolutions of $q_0$ (at $\rho_p=0.1$) in both 2D (dashed line) and 3D (solid line) simulations. (d) The saturated $q$ profiles in different cases. The blue solid line represents the 3D simulation with a saturated $q_0$ of 0.99. The orange dashed line represents the 2D simulation, with $q_0 = 0.79$ at 1827 ms (unsaturated). The green dash-dotted line represents an additional 2D simulation using an increased resistivity ($\times 100$) for obtaining the saturated $q$ profile with much less computational cost, and $q_0$ saturates at 0.75.}
    \label{figq3D}
\end{figure}

The temporal evolution of the $q$ profile obtained from the axisymmetric field of the 3D simulation is plotted in Fig. \ref{figq3D} (a). It can be seen that the safety factor in the plasma core first decreases to unity from 1.04 within 100 ms, then saturates around 0.99 over a wide radial range ($\rho_p < 0.3$) after 200 ms. Similar to Fig. \ref{figFastdynamo3D} (b), the evolution of the slow dynamo emf throughout the 3D simulation is plotted in Fig. \ref{figq3D} (b), which exhibits a negative profile in the plasma core and presents a high correlation with the evolution of the $q$ profile. The slow dynamo is more stable without oscillations, and the amplitude is three orders of magnitude smaller than the fast dynamo described above. The detailed role of the slow dynamo in clamping $q_0$ around unity and sustaining flux pumping will be discussed in Sec. \ref{secDynamo}.

The evolutions of $q_0$ (at $\rho_p=0.1$) and saturated $q$ profiles from 2D and 3D simulations are plotted in Fig. \ref{figq3D} (c) and (d) for direct comparison. It is clear that in the 3D simulation, $q_0$ is maintained around unity, while it decreases significantly below unity in the 2D simulation. Specifically, the 2D simulation with increased resistivity ($
\times 100$) shows the saturation of $q_0$ around 0.75. The significant difference of $q_0$ between the 2D and 3D simulations indicates an anomalous current redistribution mechanism due to the 3D effect in the presence of MHD instability. The evolution behaviours of $q_0$ in the 2D and 3D simulations are respectively consistent with the 'modeled' and 'experimental' curves in Fig. 3 of Ref. \cite{Burckhart2023NF}.

\begin{figure}[htb]
    \centering
    \includegraphics[width = 0.8\textwidth]{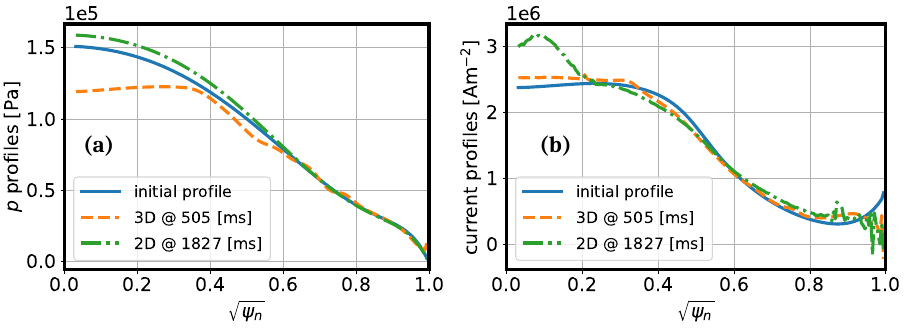}
    \caption{Radial profiles of (a) plasma pressure and (b) toroidal current density in the 2D (at 1827 ms, almost saturated) and 3D (at 505 ms, saturated) simulations.}
    \label{figpJ3D}
\end{figure}

Comparisons on detailed profiles (flux surface averaged) of plasma pressure and toroidal current density are presented in Fig. \ref{figpJ3D}. Significant differences between the 2D and 3D simulations can be seen in both pressure and current density profiles. Specifically, in the 2D simulation, the pressure gradient is maintained due to the absence of MHD instabilities. The slight increase in absolute pressure value can be attributed to the residual effect resulting from the cancellation of perpendicular heat conductivities and heating sources. Meanwhile, in Fig. \ref{figpJ3D} (b), the maximum current density in the plasma core for the 2D simulation is lifted above 3 MA/m$^{2}$ from 2.4 MA/m$^{2}$, i.e., increased by about 1/3 compared to the initial value. However, in the 3D simulation, the central profiles are flattened in a wide range. The pressure gradient in the 3D simulation almost vanishes in the plasma core, and the amplitude of the central plasma current density ( $\approx$ 2.5 MA/m$^{2}$) remains almost the same as its initial value. The flattened pressure and current density profiles in the 3D simulation together result in a low-shear $q$ profile in the plasma core, as shown in Fig. \ref{figq3D} (d), and $q_0$ is sustained approximately at unity preventing the destabilization of 1/1 internal kink mode and the subsequent sawtooth oscillations.

\begin{figure}[htbp]
    \centering
    \includegraphics[width = 0.8\textwidth]{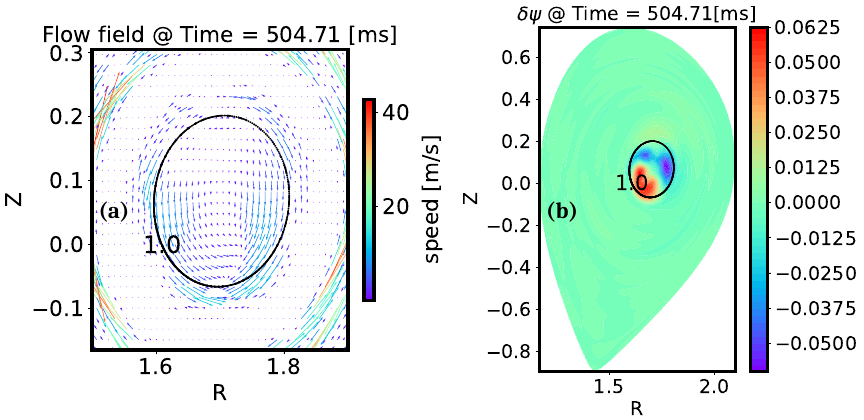}
    \caption{Non-axisymmetric components (n > 0) of (a) plasma flow and (b) poloidal magnetic flux in the 3D simulation at 505 ms. The color map in (a) represents the amplitude of the velocity speed. The black dotted line indicates the $q = 1$ rational surface.}
    \label{figModestructure3D}
\end{figure}

The non-axisymmetric components (n > 0) of mode structures in plasma flow and poloidal magnetic flux at 505 ms for the 3D simulation are plotted in Fig. \ref{figModestructure3D}. An obvious 1/1 convection flow field is still observable in Fig. \ref{figModestructure3D} (a) at the nonlinear quasi-stationary stage, but the structure is less symmetric compared with that of the linear stage of the 1/1 quasi-interchange mode as shown by Fig. \ref{fig3DEnergiesModestructureLinear} (b). The distorted convection cell in the saturated stage of the 3D simulation could be caused by the locally advected shift of the magnetic axis and the magnetic reconnection resulting from the entrance of a $q=1$ rational surface. The amplitude of the plasma velocity around the $q = 1$ rational surface is a few m/s, similar to that estimated for the flux pumping discharge (\#164661) of DIII-D in the presence of an externally induced helical core \cite{Piovesan2017NF}. However, the plasma flow speed is too small to be measured experimentally and a direct comparison against the AUG experiment is not yet available. The related dynamo effect generated by this convection field will be analyzed in Sec. \ref{secDynamo}. Besides, Fig. \ref{figModestructure3D} (b) presents the  non-axisymmetric mode structure of $\psi$, which is also dominated by a 1/1 perturbation. The results indicate that the 1/1 helical mode in the plasma core should be responsible for the flux pumping mechanism manifested by the redistribution of pressure and current density, as described above.

\begin{figure}[htbp]
    \centering
    \includegraphics[width = 0.8\textwidth]{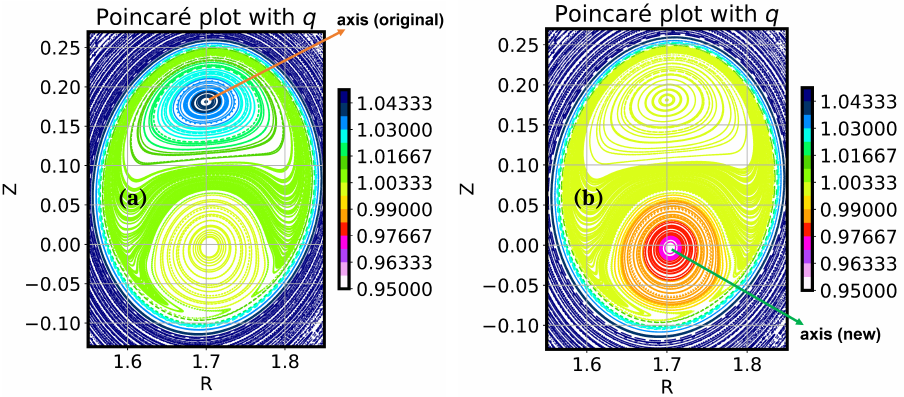}
    \caption{Poincaré plot for the 3D simulation at 505 ms (at the $\varphi=0$ plane). The color map represents the localized $q$ values calculated based on the total magnetic field but with different choices of magnetic axis: (a) the upper original axis as the reference and (b) the lower new axis as the reference, respectively.}
    \label{figPoincare3D}
\end{figure}

Fig. \ref{figPoincare3D} presents the Poincaré plots for the 3D simulation at 505 ms at $
\varphi=0$ poloidal plane. In addition to the original magnetic axis at the upper half plane, a 1/1 magnetic island-like flux tube forms at the lower half plane, which is quite similar to the 1/1 resistive tearing mode. These two magnetic flux tubes are helically twisted with respect to each other, and their global mode numbers are both m/n = 1/1. The original magnetic tube can be considered a shifted and twisted form of the initial nested magnetic surfaces around the axis, resulting from the early 1/1 ideal quasi-interchange instability. However, the new magnetic island-like flux tube should be the result of later magnetic reconnection in the presence of resistivity and a $q = 1$ rational surface. Consequently, based on their different natures (ideal vs. resistive), we can reasonably infer that the spiral behaviour of magnetic field lines in each flux tube may differ from the counterpart. To clarify this, we calculate the 3D localized $q$ by tracing the total magnetic field and considering the helical magnetic axis \cite{Zhang2020q_evolution}. The values of $q$ at each Poincaré point, with different choices of the reference magnetic axis, are indicated by the color map in Fig. \ref{figPoincare3D} (a) and (b), respectively. In the first instance, we choose the original axis (the upper at $\varphi=0$) as the reference and trace magnetic field lines over hundreds of toroidal periods to calculate the $q$ values. Fig. \ref{figPoincare3D} (a) shows that inside the original plasma core, $q_0$ remains around 1.04, almost the same as the initial value. This result indicates that the topology of the magnetic field in the original core has not changed in the 3D simulation due to the conservation of magnetic flux in the presence of ideal MHD instability. In the other instance, we choose the new axis (the lower at $\varphi=0$) as the reference to calculate $q$ values, the result is presented in Fig. \ref{figPoincare3D} (b). We observe that the local $q_0$ in the new flux tube is slightly below unity at about 0.97. In both cases, the light green region of the counterpart flux tube represents the value of exact unity, which is trivial because the global helicities of both flux tubes are m/n = 1/1. Nevertheless, the different localized $q_0$ values inside the two flux tubes indicate they have different rotational transforms. Compared with Fig. \ref{figModestructure3D} (a), the convective plasma flow is mainly localized to the X-point, which is underneath the new flux tube (the lower at $\varphi=0$), suggesting an incomplete magnetic reconnection \cite{Zhang2020NFfluxpumping}. With the flattened pressure profile shown in Fig. \ref{figpJ3D} (a), the equilibrium should be more stable for the quasi-interchange mode. Therefore, the convection flow generated by this 1/1 tearing mode should be crucial for maintaining the helical states and keeping $q_0$ around unity. The $q_0$ values calculated in different ways (with the axisymmetric magnetic field or the total magnetic field) are all close to unity, guaranteeing the suppression of sawteeth in the AUG flux pumping discharge.

\section{Slow dynamo and flux pumping on the resistive diffusion timescale} \label{secDynamo}
Previous theoretical studies revealed the critical role of the dynamo effect in the flux pumping process \cite{Jardin2015PRL, Krebs2017PoPFluxPumping}. To understand the nonlinear plasma dynamics in the 3D flux pumping simulation, in this section, the redistribution mechanisms of the slow dynamo effect on the magnetic flux and plasma current density are carefully analyzed, which lead to the helical quasi-stationary state without sawteeth as presented above. For the sake of simplicity, the term 'dynamo' in this section will be used exclusively to refer to the slow dynamo, as we are analyzing the flux pumping that occurs on the resistive diffusion timescale.

\subsection{Dynamo term in the toroidal direction and magnetic flux conservation} \label{subSectionDynamoToroidal}
The governing equation determining the evolution of the magnetic field is the induction equation (Eq. \ref{eq1induction}). In JOREK simulations, the external loop voltage is mainly applied on the plasma boundary to maintain the total plasma current. For the core plasma instabilities of interest in this study, the influence of the external loop voltage is negligible. Meanwhile, the electrostatic potential term in Eq. \ref{eq1induction} is ignored since we will focus on the axisymmetric components averaged over the magnetic flux surfaces. Following the derivation of Ref. \cite{Krebs2017PoPFluxPumping}, in cylindrical coordinates $(R, Z, \varphi)$, the induction equation along the toroidal direction for the poloidal magnetic flux $\psi$ ($\equiv RA_\varphi$) is written as
\begin{equation}\label{eqInductionPsi}
    \begin{split}
    \dfrac{\partial\psi}{\partial t} = R\left[\left(\mathbf{v}\times\mathbf{B}\right)_\varphi-\eta\left(J_\varphi-S_{j}\right)\right].
    \end{split}
\end{equation}
For the 3D simulation, we split all quantities into an axisymmetric part ($f_0$)\footnote[4]{Note that the subscript 0 on the physical quantities (except $q_0$, $\mu_0$) in this section denotes the axisymmetric component of the 3D simulation, distinguishing it from the magnetic axis values that appear above.} and a non-axisymmetric part ($f_1$), then the toroidally averaged component of Eq. \ref{eqInductionPsi} becomes
\begin{equation}\label{eqInductionPsiAvg}
    \begin{split}
    \dfrac{\partial\psi_0}{\partial t} = R\left[\left(\mathbf{v}_0\times\mathbf{B}_0\right)_\varphi+\left(\mathbf{v}_1\times\mathbf{B}_1\right)_{\varphi, n=0}-\eta_{0}\left(J_{\varphi,0}-S_{j}\right)-\left(\eta_{1}J_{\varphi,1}\right)_{n=0}\right].
    \end{split}
\end{equation}
In previous research \cite{Krebs2017PoPFluxPumping}, the assumption of $\mathbf{v}_0\simeq 0$ is made to remove the first term on the right-hand side of Eq. \ref{eqInductionPsiAvg}. However, in the present 3D nonlinear simulation with realistic parameters, the contribution to the toroidal electric field by the axisymmetric plasma flow and the magnetic field is non-ignorable, and is therefore retained in the derivation and analysis.

However, for the 2D simulation, we find the assumption of $\mathbf{v}_0
\simeq 0$ still holds, and the corresponding induction equation is as follows
\begin{equation}\label{eqInductionPsi2D}
    \begin{split}
    \dfrac{\partial\psi_{2D}}{\partial t} = -R\eta_{2D}\left(J_{\varphi,{2D}}-S_{j}\right).
    \end{split}
\end{equation}

We linearize axisymmetric parts of $\psi_0,\eta_0$ and $J_{\varphi,0}$ in Eq. \ref{eqInductionPsiAvg} with respect to their 2D components (e.g, $\eta_0=\eta_{2D}+\Delta\eta$), and subtract it by Eq. \ref{eqInductionPsi2D}. Then the induction equation for the difference in the axisymmetric components (n  = 0) of the 2D and 3D simulations can be written as
\begin{equation}\label{eqInductionPsiStationary}
    \begin{split}
    \dfrac{1}{R}\dfrac{\partial\Delta\psi}{\partial t} = \left(\mathbf{v}_0\times\mathbf{B}_0\right)_\varphi+\left(\mathbf{v}_1\times\mathbf{B}_1\right)_{\varphi, n=0}-\eta_{2D}\Delta J_\varphi-\Delta\eta \left(J_{\varphi,0}-S_{j}\right).
    \end{split}
\end{equation}
The main difference of Eq. \ref{eqInductionPsiStationary} compared to Eq. 6 in Ref. \cite{Krebs2017PoPFluxPumping} is the inclusion of $\left(\mathbf{v}_0\times\mathbf{B}_0\right)_\varphi$ and current source $S_j$ on the right-hand side. The nonlinear resistive term $
\left(\eta_1 J_{\varphi,1}\right)_{n=0}$ in Eq. \ref{eqInductionPsiAvg} is ignored here because its amplitude ($\sim 10^{-3}$ mV/m) is much smaller than the target toroidal electric field (of the order of mV/m).

\begin{figure}[thb]
    \centering
    \includegraphics[width = 0.8\textwidth]{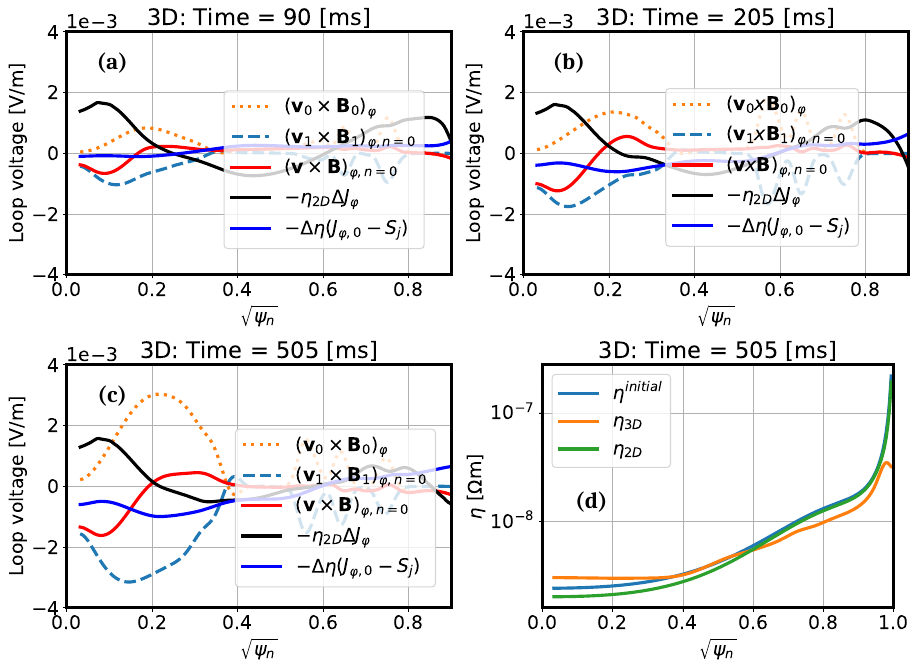}
    \caption{(a-c) Radial distributions of different flux surface averaged terms contributing to the axisymmetric toroidal electric field on the right-hand side of Eq. \ref{eqInductionPsiStationary}, respectively at 90 ms, 205 ms, and 505 ms of the 3D simulation. (d) Radial distributions of resistivity for the initial equilibrium (blue line), the time point of 505 ms in the 3D simulation (orange line), and the saturated profile in the 2D simulation (green line). Note that the saturated results in the 2D simulation, including the resistivity profile and plasma current density, are obtained by increasing the resistivity by 100 times. When plotting together with the 3D result, the resistivity in the 2D simulation is scaled down by 0.01 to represent the realistic situation.}
    \label{figBalanceConditions}
\end{figure}

The various contributions by different terms on the right-hand side of Eq. \ref{eqInductionPsiStationary} are illustrated in Fig. \ref{figBalanceConditions} for different moments. In all three stages of the 3D simulation, as shown in Fig. \ref{figBalanceConditions}, the black solid lines show almost the same electric field deficit $\eta_{2D}\Delta J_\varphi$ of $-$1.6 mV/m. This deficit corresponds to the required electric field that should be provided by an anomalous current diffusion mechanism to maintain the quasi-stationary flat current density profile in the plasma core of the 3D simulation, as shown by Fig. \ref{figpJ3D} (b). The blue solid lines indicate the negative electric field contributed by the resistivity flattening. The contribution of the resistivity flattening first increases slowly over time and then saturates. This is because as implied by the orange dotted line in Fig. \ref{figpJ3D} (a), the flux pumping mechanism significantly redistributes the electron temperature in the plasma core and results in a flattened and increased resistivity in the plasma core of the 3D simulation. Besides the resistivity flattening, the dominant mechanism offsetting the driving effect of the current source originates from the nonlinear dynamo term from the non-axisymmetric components, i.e., the 1/1 MHD instability in this study. Specifically, a clear profile of the negative dynamo effect $\left(\mathbf{v}_1\times\mathbf{B}_1\right)_{\varphi, n=0}$ forms in the plasma core and increases over time, as shown by the blue dashed lines in Fig. \ref{figBalanceConditions}. Nevertheless, we also observed a finite positive contribution on the toroidal electric field in the plasma core from the axisymmetric component of plasma flow, i.e., $\left(\mathbf{v}_0\times\mathbf{B}_0\right)_{\varphi}$. It offsets a portion of the negative dynamo from the instability, and eventually results in a discounted negative electric field, which will be referred as the net dynamo, labeled by $\left(\mathbf{v}\times\mathbf{B}\right)_{\varphi, n = 0}$ and red solid lines in Fig. \ref{figBalanceConditions}. 

\begin{figure}[htbp]
	\centering
    \includegraphics[width=0.5\textwidth]{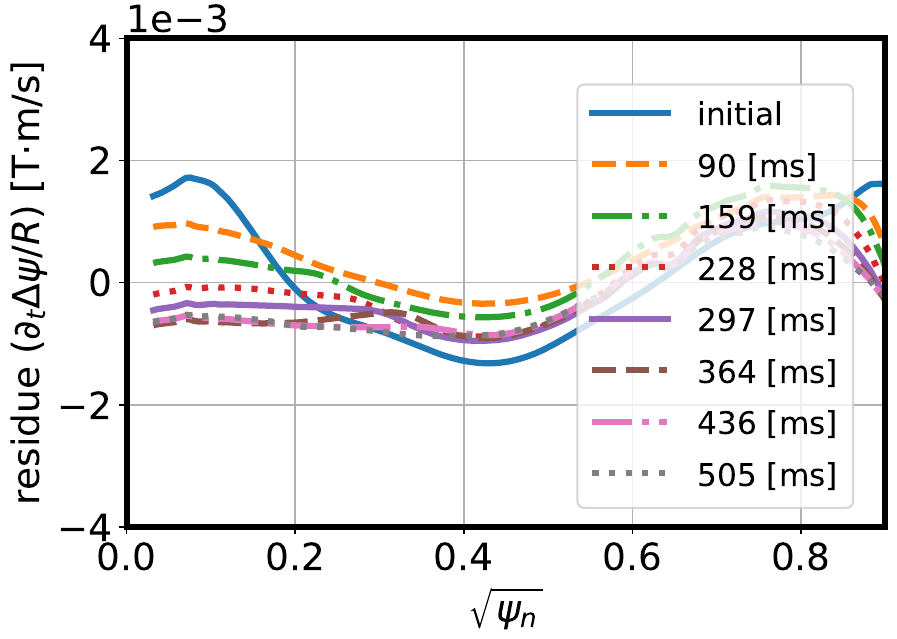}
	\caption{The residue profiles obtained by summing up all flux surface averaged terms on the right-hand side of Eq. \ref{eqInductionPsiStationary} at different moments of the 3D simulation, which are equivalent to the amplitudes of $\left(\partial_t\Delta\psi\right)/R$.}
	\label{figResidue}
\end{figure}

At the early time point of 90 ms, as shown in Fig. \ref{figBalanceConditions} (a), the resistivity flattening effect is ignorable, and the net dynamo amplitude is about $-$0.3 mV/m in the plasma core, much weaker than the required electric field deficit of $-$1.6 mV/m. As a result, $q_0$ decreases over time as shown in Fig. \ref{figq3D} (c). At 205 ms as Fig. \ref{figBalanceConditions} (b) shows, the resistivity flattening contribution is $-$0.25 mV/m, while the net dynamo is $-$1.35 mV/m. Therefore, these two effects play together to balance the electric field deficit generated by the external current drive and result in the saturation of $q_0$ as shown in Fig. \ref{figq3D} (c). At 505 ms of the 3D simulation, the resistivity flattening contribution is $-$0.3 mV/m, while the net dynamo is $-$1.6 mV/m. The sum of these two negative contributions is slightly larger than the required electric field deficit, which seems should destroy the quasi-stationary state of the 3D simulation. However, the residues of Eq. \ref{eqInductionPsiStationary} [equivalent to $\left(\partial_t\Delta\psi\right)/R$] for different moments are plotted in Fig. \ref{figResidue}. After 363 ms, we find the residual electric field is almost unaltered until the end of the 3D simulation, exhibiting a remarkably flat profile across a considerable radial range ($\rho_p<0.3$). The flattened distribution of the residual electric field corresponds to a constant shift of $\psi$ over time in the plasma core, but it will not significantly change the helicity of magnetic field lines and the profile of current density, which are mainly determined by the first or second-order spatial derivatives of $\psi$. On the other hand, not shown here, detailed analyses on the evolution of toroidal magnetic field also demonstrate the conservation of the toroidal magnetic flux $\psi_t$ at the quasi-stationary stage of the 3D simulation, where a similar cancellation between the two nonlinear coupling terms of [$\mathbf{v}_{0}$, $\mathbf{B}_0$] and [$\mathbf{v}_{1}$, $\mathbf{B}_1$] is observed in the induction equation of $B_\varphi$. Note that the resistivity is only applied in the toroidal direction of Eq. \ref{eq1induction}, thereby restricting the evolutions of $A_R$, $A_Z$, and $B_\varphi$ within the ideal MHD framework. It can thus be confirmed that the saturation of $q_0$, as observed in the 3D simulation, is a consequence of the settled radial profile of poloidal magnetic flux and the conservation of the toroidal magnetic flux.

The amplitude of the toroidal dynamo electric field obtained in the JOREK 3D MHD simulation is of the order of mV/m, which is consistent with the experimentally reconstructed electric field deficit for flux pumping discharges in AUG \cite{Burckhart2023NF} and DIII-D \cite{Piovesan2017NF}. Previous simulations usually adopt increased resistivity to reduce the computational time. However, such treatment on resistivity in 3D simulations leads to the significant overestimation of the toroidal dynamo electric field, which could be of the order of 0.1 V/m \cite{Burckhart2023NF, Krebs2017PoPFluxPumping}. The 2D and 3D MHD simulations with fully realistic parameters presented here provide the first quantitative result on the flux pumping and dynamo effect in tokamak hybrid scenarios. Nevertheless, the specific values of the toroidal dynamo electric field could be different between the MHD simulation and the experiment reconstruction due to different amplitudes and profiles of the current source used in these modelling methodologies. For this reason, a preliminary parameter scan is performed in Sec. \ref{secScan} to investigate the influence of the strength and profile of the current sources.

\subsection{The parallel dynamo emf projected to the mean magnetic field} \label{subSectionDynamoParallel}
In the above discussions, the two separated terms [$\left(\mathbf{v}_0\times\mathbf{B}_0\right)_\varphi$ and $\left(\mathbf{v}_1\times\mathbf{B}_1\right)_{\varphi, n=0}$] contributing to the negative net dynamo are actually the natural consequence of choosing the toroidal direction as the reference when analyzing the balance condition, as shown by Eqs. \ref{eqInductionPsi}-\ref{eqInductionPsiStationary}. In this subsection, we will show that the negative net dynamo, presented by the red solid lines in Fig. \ref{figBalanceConditions}, is equivalent to the dynamo contributed by the n $\ge$ 1 components of plasma flow and magnetic field if we choose the n = 0 mean magnetic field as the reference, rather than selecting the toroidal direction. Here we will refer to the obtained dynamo as the \textbf{parallel dynamo emf}. It is found solely contributed by the non-axisymmetric MHD instabilities and is therefore more consistent with the original definition of the dynamo, e.g., \textit{'the fluctuation-induced emf along the mean magnetic field'} \cite{JI2001alphaDynamoRFP}.

The parallel dynamo emf in the 3D simulation can be obtained by projecting the total induction equation (Eq. \ref{eq1induction}) along the axisymmetric magnetic field $\mathbf{B}_0$ (which we call the mean field), that is
\begin{equation} \label{eqInductionParallel}
    \begin{split}
        \mathbf{B}_0\cdot\dfrac{\partial\mathbf{A}}{\partial t}&=\mathbf{B}_0\cdot\left[\mathbf{v}\times\mathbf{B}-\eta\left({J}_\varphi-{S}_j\right)\hat{e}_\varphi\deleted{-\nabla\Phi}\right] \\
        &=\mathbf{B}_0\cdot\left[\left(\mathbf{v}_0+\mathbf{v}_1\right)\times\mathbf{B}_1-\eta\left({J}_\varphi-{S}_j\right)\hat{e}_\varphi\right].
    \end{split}
\end{equation}
Same as the derivation in Subsec. \ref{subSectionDynamoToroidal}, the externally applied boundary loop voltage and electrostatic potential are neglected, and the physical quantities are divided into their respective axisymmetric ($f_0$) and non-axisymmetric ($f_1$) components. In Eq. \ref{eqInductionParallel}, the contribution by the axisymmetric plasma flow, i.e., $\mathbf{v}_0\times\mathbf{B}_0$, is naturally excluded since it is perpendicular to the mean magnetic field. Then the toroidally axisymmetric component of Eq. \ref{eqInductionParallel} can be rewritten as
\begin{equation}\label{eqInductionParallel2}
    \begin{split}
        \bar{\mathbf{b}}_0\cdot\dfrac{\partial\mathbf{A}_0}{\partial t} =&\bar{\mathbf{b}}_0\cdot\left\{\deleted{\mathbf{v}_0\times\mathbf{B}_1}+\left(\mathbf{v}_1\times\mathbf{B}_1\right)_{n=0}\right\} \\
        &\bar{\mathbf{b}}_0\cdot\left\{-\left[\eta_0\left(J_{\varphi,0}-S_j\right)\deleted{+\eta_1\left( J_{\varphi,0} -S_j \right)+\eta_0 J_{\varphi,1}+\left(\eta_1 J_{\varphi,1}\right)_{n=0}}\right]\hat{e}_\varphi\right\} \\
        =&\bar{\mathbf{b}}_0\cdot\left[\left(\mathbf{v}_1\times\mathbf{B}_1\right)_{n=0}-\eta_0\left(J_{\varphi,0}-S_j\right)\hat{e}_\varphi\right],
    \end{split}
\end{equation}
where $\bar{\mathbf{b}}_0=\mathbf{B}_0/B_{\varphi, \text{eq}}$ (note that $B_{\varphi, \text{eq}}$ is used for normalization because it has lower numerical noise). The linear terms in Eq. \ref{eqInductionParallel2} are eliminated since we are concerned about the axisymmetric components contributing to the dynamo. The nonlinear resistive current diffusion term $\left(\eta_1 J_{\varphi,1}\right)_{n=0}$ is negligible as mentioned above. The first term in the last line of Eq. \ref{eqInductionParallel2} shows the parallel dynamo emf, which is exclusively contributed by the correlated perturbations of plasma flow and magnetic field from the non-axisymmetric MHD instabilities.

\begin{figure}[htbp]
	\centering
    \includegraphics[width=0.5\textwidth]{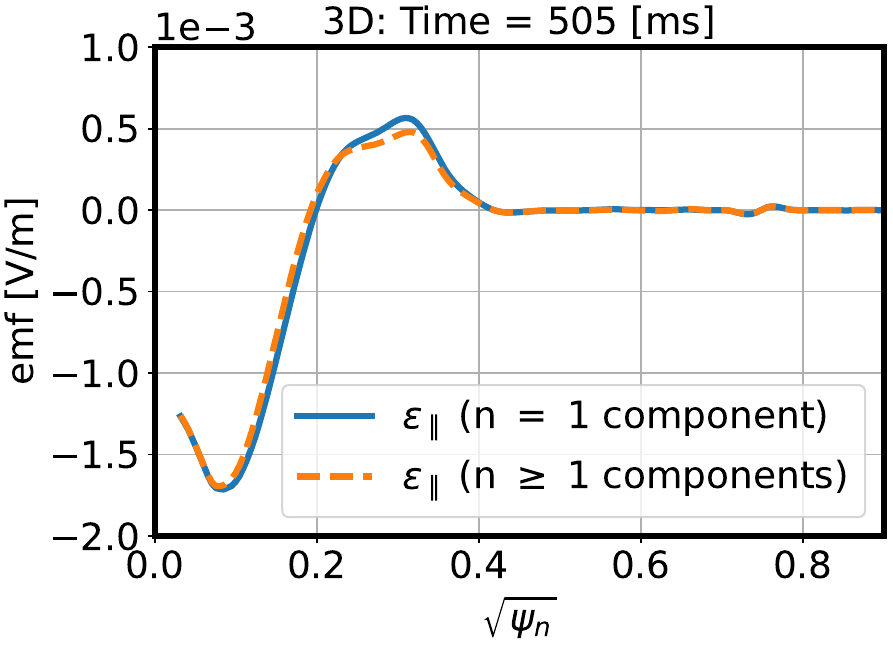}
	\caption{The parallel emf along the mean magnetic field generated by the parallel dynamo emf at 505 ms of the 3D simulation. The solid line corresponds to the dynamo term calculated only with the n = 1 component, and the dashed line is from all n $\ge$ 1 components.}
	\label{figTrueDynamoEmf}
\end{figure}

The same approach as in Subsec. \ref{subSectionDynamoToroidal} can be applied to Eq. \ref{eqInductionParallel2} to linearly expand the axisymmetric components ($f_0$) to the values of the 2D simulation ($f_{2D}$), so that we can obtain similar profiles of the electric field deficit and resistivity flattening term. However, in this subsection, we will focus on the amplitude of the parallel dynamo emf that is averaged along the magnetic flux surfac and contributed by the non-axisymmetric MHD instabilities, i.e., $\varepsilon_\parallel=\langle\bar{\mathbf{b}}_0\cdot\left(\mathbf{v}_1\times\mathbf{B}_1\right)_{n=0}\rangle$. The radial distribution of $\varepsilon_\parallel$ is plotted in Fig. \ref{figTrueDynamoEmf}. The parallel dynamo emf is contributed dominantly by the n = 1 MHD instability. Its profile and amplitude are almost the same as the net dynamo shown by the red solid line in Fig. \ref{figBalanceConditions} (c). 

The two methodologies presented in Subsecs. \ref{subSectionDynamoToroidal} and \ref{subSectionDynamoParallel} are in principle equivalent with each other. However, by analyzing the problem along the mean magnetic field lines, we avoid the positive term associated with the axisymmetric plasma flow in Fig. \ref{figBalanceConditions}, thereby obtaining the parallel dynamo emf exclusively from the non-axisymmetric components of the MHD instabilities. The former concept, i.e., the net dynamo in the toroidal direction, is more intuitive and useful for tokamak plasmas, especially if one is interested in the evolution or redistribution of the poloidal magnetic flux and toroidal current density. The latter concept of the parallel dynamo emf along the mean magnetic field line is more readily comprehensible concerning the self-regulation process of plasma. This process is initiated by the helical distortion of the magnetized plasma by non-axisymmetric MHD instabilities, which generates a parallel emf to resist the resistive diffusion or current drive and sustain the mean magnetic field.

\subsection{Current redistribution mechanism from the dynamo effect} \label{subsecCurrentDiffusion}
This subsection presents a further analysis of the current redistribution mechanism resulting from the dynamo effect in the 3D simulation, which acts as an enhanced current diffusion, competing with the driving effect of the current source in the plasma core. The temporal evolution equation of toroidal current density can be derived by taking the time derivative on Ampere's law and substituting the induction equation Eq.\ref{eq1induction} into it. After some algebra, we obtain the following current density evolution equation
\begin{equation}\label{eqJphievolution}
    \begin{split}
    \dfrac{\partial J_\varphi}{\partial t}&=\hat{e}_\varphi\cdot\dfrac{1}{\mu_0}\nabla\times\left(\nabla\times\dfrac{\partial\mathbf{A}}{\partial t}+\dfrac{1}{R}\dfrac{\partial F}{\partial t}\hat{e}_\varphi\right)\\
    & = \dfrac{1}{\mu_0R}\dfrac{\partial}{\partial\varphi}\left[\nabla\cdot\left(\mathbf{v}\times\mathbf{B}-\eta\left({J}_\varphi-{S}_j\right)\hat{e}_\varphi\right)\right]-\dfrac{1}{\mu_0}\left[\nabla^2\left(\mathbf{v}\times\mathbf{B}-\eta\left({J}_\varphi-{S}_j\right)\hat{e}_\varphi\right)\right]_\varphi.
    \end{split}
\end{equation}
Apply magnetic surface average $\langle \cdots \rangle$ on Eq. \ref{eqJphievolution} and ignore the toroidal derivative terms, then Eq. \ref{eqJphievolution} becomes
\begin{equation}\label{eqJphiAvgevolution}
    \begin{split}
    \dfrac{\partial\langle J_\varphi\rangle}{\partial t} & = -\dfrac{1}{\mu_0}\langle\nabla^2\left[\mathbf{v}\times\mathbf{B}-\eta\left({J}_\varphi-{S}_j\right)\hat{e}_\varphi\right]\rangle_\varphi \\
    & = -\dfrac{1}{\mu_0}\langle\nabla^2\left[\mathbf{v}\times\mathbf{B}-\eta\left({J}_\varphi-{S}_j\right)\hat{e}_\varphi\right]_\varphi\deleted{-\dfrac{1}{R^2}\left[\mathbf{v}\times\mathbf{B}-\eta\left({J}_\varphi-{S}_j\right)\hat{e}_\varphi\right]_\varphi}\rangle \\
    & \simeq \dfrac{1}{\mu_0}\left(\left|\nabla\psi\right|^2\dfrac{\partial^2}{\partial\psi^2}+\nabla^2\psi\dfrac{\partial}{\partial\psi}\right)\langle-\left(\mathbf{v}\times\mathbf{B}\right)_\varphi+\eta\left(J_\varphi-S_j\right)\rangle.
    \end{split}
\end{equation}
In the last step of Eq. \ref{eqJphiAvgevolution}, when transforming from the cylindrical coordinates $\left(R,Z,\varphi\right)$ to the general field-aligned toroidal coordinates $\left(\psi,\theta,\varphi\right)$, the derivatives over the generalized poloidal angle $\theta$ are ignored as well because we focus on the m/n = 0/0 contributions of dynamo and resistive terms. The blue term in Eq. \ref{eqJphiAvgevolution} proportional to $1/R^2$ is ignored because it is much smaller than the first term on the right-hand side, i.e., $R^{-2}/\nabla^2\sim\mathcal{O}\left(L^2/R^2\right)\ll 1$, where $L$ is the characteristic spatial scale of the 1/1 mode and is of the order of 0.1 m.

\begin{figure}[htbp]
    \centering
    \includegraphics[width = 0.8\textwidth]{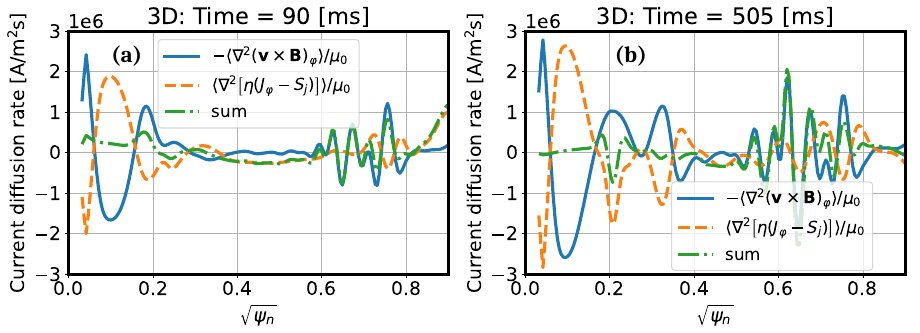}
    \caption{The radial profiles of the two terms contributing to the evolution of flux surface averaged toroidal plasma current density in the 3D simulation at (a) 90 ms and (b) 505 ms, as shown by the right-hand side of Eq. \ref{eqJphiAvgevolution}. The solid line indicates the contribution of the net dynamo effect to the current diffusion, the dashed line indicates the net current drive in the presence of the current source and resistive current diffusion, and the dash-dotted line represents the sum of the two terms, which corresponds to the total current diffusion rate.}
    \label{figBalanceCurrentConditions}
\end{figure}

The radial distributions of two contributing terms at two different moments in the 3D simulation are plotted in Fig. \ref{figBalanceCurrentConditions}. The outer structures ($\rho_p>0.5$) contributed by some tearing instabilities are not the subject of our study, as our primary focus is on the dynamo induced current redistribution within the core region ($\rho_p<0.4$).  The blue solid line indicates the net dynamo term, i.e., $-\langle\nabla^2\left(\mathbf{v}\times\mathbf{B}\right)_\varphi\rangle/\mu_0$. The orange dashed line corresponds to the net current driving effect in the presence of current source and resistive current diffusion, i.e., $\langle\nabla^2\left[\eta\left(J_\varphi - S_j\right)\right]\rangle/\mu_0$. The green dash-dotted line plots the sum of these two terms, corresponding to the total current diffusion rate. As shown by Fig. \ref{figBalanceCurrentConditions} (a), at 90 ms, the positive net current drive is slightly larger than the negative diffusion effect from the net dynamo term in the plasma core, which is consistent with the slow decline of $q_0$ in Fig. \ref{figq3D} (c). Nevertheless, at 505 ms, the excellent cancellation between these two terms is achieved, as shown in Fig. \ref{figBalanceCurrentConditions} (b), corresponding to the saturation of current density and $q_0$ in the plasma core. In the 2D simulation, the plasma flow $\mathbf{v}$ is ignorable. Therefore the positive driving effect from the current source term cannot be balanced by the dynamo, resulting in a continuous growth of current density in the plasma core region before the final saturation. Meanwhile, the smoothed current source used in the simulation will result in a less intensive driving effect in the current diffusion equation, since the related term in Eq. \ref{eqJphiAvgevolution} is roughly proportional to the second derivative of the current source profile.

The analyses in Figs. \ref{figBalanceConditions} and \ref{figBalanceCurrentConditions} both demonstrate the role of negative net dynamo term in counteracting the driving effect from the current source. The results are consistent with the experimental observation where the sawtooth oscillations are avoided by keeping $q_0$ around unity on the timescale of resistive current diffusion.

\section{Preliminary parameter scan of the current source}\label{secScan}

\begin{figure}[htbp]
	\centering
    \includegraphics[width=0.5\textwidth]{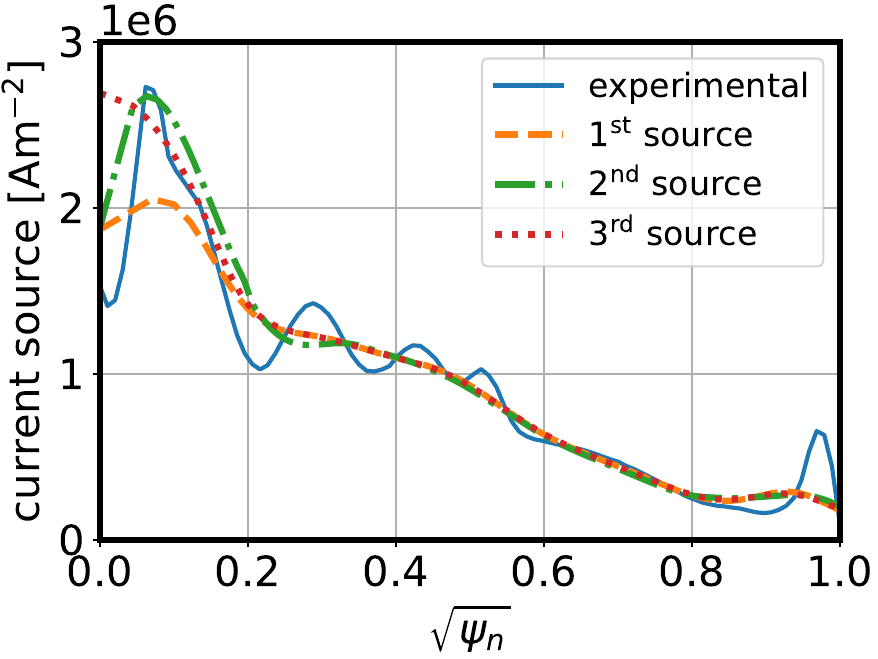}
	\caption{Radial profiles of different non-inductive current sources: the unmodified experimental current source (solid line), the first current source (dashed line), the second current source (dash-dotted line), the third current source (dotted line).}
	\label{figCurrentSources}
\end{figure}

\begin{figure}[htbp]
    \centering
    \includegraphics[width = \textwidth]{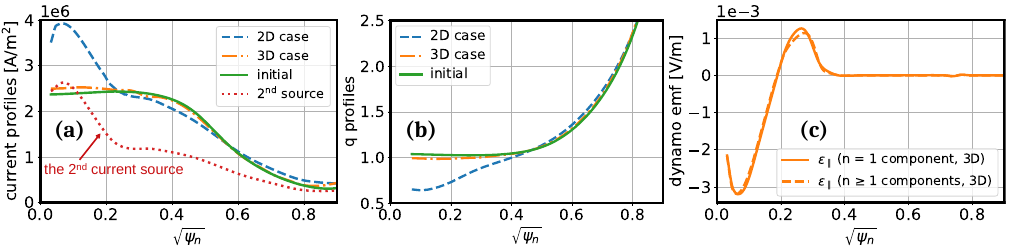}
    \caption{Simulation results with the second current source [dotted line in panel (a)]: (a) current density and (b) $q$ profiles at the initial stage (solid line), the saturated stages of 2D (dashed line) and 3D (dash-dotted line) simulations; and (c) parallel dynamo emf along the mean magnetic field in the 3D simulation, calculated using n = 1 component (solid line) and all n $\geq$ 1 components (dashed line), respectively.}
    \label{fig2ndTypeSource}
\end{figure}

\begin{figure}[htbp]
    \centering
    \includegraphics[width = \textwidth]{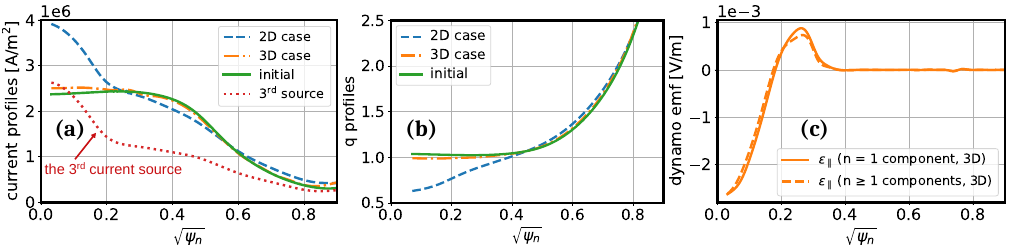}
    \caption{Simulation results with the third current source [dotted line in panel (a)]: (a) current density and (b) $q$ profiles at the initial stage (solid line), the saturated stages of 2D (dashed line) and 3D (dash-dotted line) simulations; and (c) parallel dynamo emf along the mean magnetic field in the 3D simulation, calculated using n = 1 component (solid line) and all n $\geq$ 1 components (dashed line), respectively.}
    \label{fig3rdTypeSource}
\end{figure}

In addition to the first current source used in the simulations above [the dashed line in Fig. \ref{figCurrentSources} or Fig. \ref{figsource} (a)], a preliminary parameter scan of the current source has been conducted to study the influence of its intensity and profile on flux pumping. Specifically, two stronger current sources are designed, as indicated by the dash-dotted and dotted lines in Fig. \ref{figCurrentSources}, with maximum intensities ($\sim 2.70$ MA/m$^2$) comparable to the experimental current source. The second current source peaks off-axis at $\rho_p = 0.1$, similar to the slightly off-axis deposition of ECCD in the experiment. The third current source peaks on the magnetic axis to simulate the ideal situation of on-axis ECCD deposition, as more central ECCD deposition results in higher current drive efficiency in experiments (approximately proportional to $T_e/n_e$) \cite{Burckhart2023NF}. The initial equilibrium and other parameters remain the same as Sec. \ref{secSetup}. However, the resistivity in 2D simulations is increased by two orders of magnitude to reduce the computational time. The 3D simulations are run for over 200 ms until $q_0$ ceases to decrease. The saturated profiles of current density, $q$, and parallel dynamo emf  are plotted in Figs. \ref{fig2ndTypeSource} and \ref{fig3rdTypeSource}, respectively for the second and third current sources.

With the second current source, the 2D simulation yields an off-axis peaked current density and a minimum $q_0\approx 0.6$ at $\rho_p = 0.1$, as shown by dashed lines in Fig. \ref{fig2ndTypeSource} (a) and (b). In contrast, the current density and $q$ profiles from the 3D simulation remain flat in the plasma core to avoid sawtooth onset due to the flux pumping mechanism, as shown by dash-dotted lines in Fig. \ref{fig2ndTypeSource} (a) and (b). Specifically, the maximum current density from the 3D simulation is significantly lower than that of the 2D simulation (2.5 vs. 4 MA/m$^{2}$), and $q_0$ remains clamped close to unity. The parallel dynamo emf is mainly generated by the n = 1 MHD mode in the 3D simulation, as shown by Fig. \ref{fig2ndTypeSource} (c). It exhibits a similar profile to the first current source case (Fig. \ref{figTrueDynamoEmf}), but with a much larger amplitude. The stronger current drive is balanced by the  larger negative dynamo, similar to the situation in Sec. \ref{subsecCurrentDiffusion}, though the details are not shown here.

With the third current source that is on-axis peaked, the 2D simulation results in a maximum current density and minimum $q_0$ ($\approx 0.6$) on the magnetic axis, as shown by dashed lines in Fig. \ref{fig3rdTypeSource} (a) and (b). The 3D simulation results demonstrate that, even if the current source (e.g., ECCD) is deposited on the magnetic axis to achieve higher efficiency, flux pumping can still prevent the accumulation of core current density and the drop of $q_0$, thereby maintaining a sawtooth-free state. As plotted by Fig. \ref{fig3rdTypeSource} (c), the parallel dynamo emf generated by the n = 1 MHD mode also exhibits an on-axis peaked distribution, counteracting the on-axis deposited current drive. 

The simulation results with the second and third current sources are in better agreement with the reconstructed profiles of current density and $q$ for the flux pumping phase of AUG discharge \#36663, (Figs. 3 and 4 in Ref. \cite{Burckhart2023NF}), despite minor differences in specific values due to the different exact time points between this study and Ref. \cite{Burckhart2023NF} (3.75 - 3.95 s vs. 4.8 s).

The preliminary scan of different current sources demonstrates the robustness of flux pumping to a reasonable extent at typical parameters of AUG hybrid scenarios. The control of the current source deposition location is primarily driven by the need for higher drive efficiency with the on-axis ECCD. However, it is not essential for sawtooth suppression, as the self-regulating process of the 1/1 MHD mode can generate the required distribution and amplitude of the dynamo emf to counteract the current peaking tendency in the plasma core. Nevertheless, a more systematic scan of the current source intensity is needed to compare the ECCD operation window for flux pumping in simulations with AUG experiments, which is beyond the scope of the present paper and will be addressed in a future publication.

\section{Conclusion and discussion} \label{secSummary}
In summary, a quantitative simulation study for the flux pumping of the hybrid scenario observed in the AUG experiments has been carried out with JOREK. We adopt the two-temperature, visco-resistive, full MHD model and choose the experimentally relevant parameters, including the Spitzer resistivity, viscosity, and anisotropic heat conductivities. The comparisons between axisymmetric 2D and non-axisymmetric 3D simulations on the resistive diffusion timescale of seconds demonstrate the effectiveness of the dynamo effect in the redistribution of plasma current density and pressure, thereby sustaining $q_0$ around unity and suppressing the sawtooth.

\subsection{Main conclusions of the present work}

The 2D simulation results show a continuous decrease of $q_0$ in the plasma core, because the non-inductive current source plays a driving role in the accumulation of core current density. The saturation amplitude of the maximum toroidal current density in the 2D simulation is about 3.2 to 4 MA/m$^2$ (depending on the strength of the current source), much larger than the initial value of 2.4 MA/m$^2$. The resistivity has limited influence on the final saturated solution of the 2D simulation, since in the absence of MHD instabilities, the equilibrium evolution is mainly determined by the resistive current diffusion process. The final saturated $q_0$ values in 2D simulations are much lower than unity (ranging from approximately 0.6 to 0.75). In principle, the $q_0$ below unity predicted by the 2D simulation should result in the onset of sawteeth, which contradicts the experimental observation of sawtooth-free phases in AUG. Therefore, the 2D model is not able to correctly describe the current diffusion process in the flux pumping discharge.

In contrast, 3D simulations with realistic resistivity and other relevant parameters first show a fast and under-damped oscillation of $q_0$ due to the fast dynamo effect during the initial saturation stage of the 1/1 quasi-interchange. The fast dynamo effect generates the toroidal electric field with a large amplitude on the order of V/m and modifies the helicity of the mean magnetic field (manifested as the change in $q_0$) on a timescale of a few milliseconds, which is much shorter than that of the resistive diffusion (on the order of seconds). Subsequent nonlinear 3D simulations on the resistive diffusion timescale show that $q_0$ first decreases slightly below unity in the first 200 ms. After that, the system enters into a quasi-stationary stage with $q_0$ saturating around 0.99. The maximum toroidal plasma current density ($\simeq$ 2.5 MA/m$^2$) is maintained close to the equilibrium value with a broad and flat profile for hundreds of milliseconds. The significant difference in the toroidal plasma current density between the 2D and 3D simulations indicates the anomalous current redistribution mechanism in flux pumping, which is equivalent to an additional toroidal electric field deficit of a few mV/m to clamp the current density in the 3D simulation. Detailed analyses show that in the quasi-stationary stage after 200 ms, the slow dynamo effect generated by the 1/1 MHD instability has a comparable but negative amplitude (a few mV/m) in the plasma core, while a limited contribution from the resistivity flattening effect is also observed. The sum of the toroidal electric fields due to the negative dynamo and resistivity flattening is comparable with or even slightly larger than the required electric field deficit. The slight mismatch results in a slow shift of $\psi$ over time but will not obviously change the helicity of the magnetic field and the value of $q_0$. Because the residue electric field is kept constant in both space and time in the plasma core, and the toroidal magnetic flux conservation is also satisfied in the quasi-stationary stage. On the other hand, the excellent cancellation between the current driving effect from the non-inductive source and the enhanced current diffusion by the slow dynamo effect is confirmed in the 3D simulation, which further validates the effective role of the dynamo in the current redistribution of flux pumping.

The preliminary scan of different current sources (on-axis or off-axis deposition) in the experimental intensity range demonstrates that the flux pumping has good robustness due to the self-regulation of the dynamo effect and the 1/1 MHD mode, which can generate the required negative electric field to pump the current-driven injected poloidal magnetic flux and current density outward from the core. The results further reinforce the previous hypothesis of maximizing the current drive efficiency by depositing ECCD in the plasma center while preventing sawtooth onset through the flux pumping induced plasma current redistribution \cite{Burckhart2023NF}.

The nature of the 1/1 continuous instability observed in the flux pumping phases of the AUG experiment cannot be directly identified based on the diagnostic data \cite{Burckhart2023NF}. In the 3D JOREK simulation, the 1/1 MHD mode initially presents the feature of the 1/1 quasi-interchange mode. Then the 1/1 magnetic island is observed in the nonlinear saturated stage. Therefore, the plasma core consists of two twisted 1/1 magnetic flux tubes. One of them is the original plasma core with the local $q_0$ remaining at the initial value of the magnetic axis ($\simeq$ 1.04), corresponding to the ideal MHD perturbation generated by the 1/1 quasi-interchange mode. The other magnetic flux tube seems to result from magnetic reconnection due to the inclusion of the $q$ = 1 rational surface and finite resistivity. The local $q_0$ in this magnetic island is slightly lower than unity ($\simeq$ 0.97), which suggests that the resistive diffusion mechanism dominates here. Consequently, the 1/1 MHD mode observed in this 3D simulation is a combination of the 1/1 ideal quasi-interchange mode and the 1/1 resistive tearing mode. However, this result is not the final conclusion, as the mode properties are quite sensitive to the value of $q_0$. Varying the current source intensity or other diffusion parameters may result in a slightly different value of saturated $q_0$ around unity. For example, if $q_0$ is maintained above unity, the 1/1 magnetic island will be avoided, and the helical core should be dominated by the 1/1 quasi-interchange mode. Future parameter scans will address this potential bifurcation behaviour of the 1/1 MHD instabilities. Nevertheless, the slow dynamo provided by the MHD instabilities will not significantly deviate from the magnitude of mV/m, which is roughly determined by the value of the resistivity and the difference in current density between the 2D and 3D simulations.

\subsection{Outlook for flux pumping modelling} \label{subsecOutlook}

In the quasi-stationary stage, the amplitude of the toroidal dynamo electric field in the JOREK simulation is of the order of mV/m, and the convective plasma flow speed is a few m/s, both are quantitatively consistent with experimental observations or estimations \cite{Burckhart2023NF, Piovesan2017NF}. However, the radial profile and amplitude of the non-inductive current source will significantly influence the evolution of toroidal plasma current density and the required deficit of electric field. The presented simulations adopt three moderately peaked current sources due to the consideration of numerical stability and the limitation of computational resources. The next step will be to further scan the values and profiles of the current source and the equilibrium pressure to roughly identify the operation window for flux pumping in AUG. We will also further compare the simulations with the existing or upcoming experimental results. 

From the standpoint of physical understanding, parameter scans of viscosity and resistivity will help build up a systematic picture of flux pumping and the dynamo effect in tokamak hybrid scenarios. For example, previous visco-resistive MHD modelling of RFP plasma has shown the bifurcation into the multiple helicity (MH) and single helicity (SH) states at high ($\gsim 10^4$) and low ($\lsim10^3$) Hartmann numbers, respectively \cite{Cappello2000PRL_RFP, Bonfiglio2005prlRFP, Cappello2006PoPRFP}. Their Hartmann numbers are much lower than the present tokamak case ($\sim 10^7$) due to the much larger resistivity and viscosity adopted in RFP modelling. Nevertheless, the quasi-stationary single helical core is still obtained in the presented tokamak modelling at AUG parameters. A brief parameter scan of the Hartmann number by increasing viscosity and resistivity has been carried out (not shown here). It is found that with the increase of Hartmann number, the quasi-stationary flux pumping state first transits into periodic sawtooth-like oscillations (H $\sim 10^5$), with $q_0$ repeatedly oscillating between 0.95 and 1.0. When the Hartmann number is further increased above $\sim 10^4$, the 3D simulations saturate with the quasi-stationary 1/1 resistive internal kink mode, where $q_0$ is much lower than unity, e.g., $\sim 0.7$. 
The dependencies of the different helical states of tokamak plasmas on viscosity and resistivity are under further investigation and will be reported separately in the near future. Viscosity is a relatively free parameter in MHD simulations and is chosen empirically at present. However, its value may significantly influence the amplitude of plasma flow, and subsequently affect the dynamo strength and the success or failure of flux pumping \cite{Shen2018NF,Zhang2020NFfluxpumping}. Whilst the resistivity can be directly determined from the experimental electron temperature, the amplitude of the dynamo electric field will scale itself spontaneously with respect to different resistivity values to balance the current driving effect, which poses an interesting question on the upper-limit of resistivity for flux pumping remaining effective.

The present MHD simulations still adopt the single fluid, visco-resistive, full MHD equations. The development of the extended MHD model for JOREK will be carried out to introduce other possible dynamo mechanisms, such as the Hall dynamo and diamagnetic dynamo \cite{JI2001alphaDynamoRFP, King2011PoPRFP, Mao2023PRR}, and also to capture more non-ideal corrections to the 1/1 MHD instabilities, such as the two-fluids effect \cite{Halpern2011PoPSawtooth, Zhang2020Hall}, finite Larmor radius (FLR) effect \cite{King2011PoPRFP}, self-consistent bootstrap current evolution \cite{Yu2024NFfp}, etc. Ideally, the helical plasma core is expected to self-regulate the amplitudes of various dynamo effects provided by the 1/1 instability in the presence of these corrections to sustain a similar quasi-stationary state with $q_0\simeq 1$. \revisioncolor{In addition to MHD simulations, the MHD-kinetic hybrid model of JOREK \cite{Bogaarts2022POP_EP_JOREK, Bergström2025PPCFrunaway} will be introduced in future flux pumping research to investigate the synergistic effects between energetic particles and the 1/1 MHD instabilities, including the impact of energetic ions and electrons \cite{Kolesnichenko2007popEP_QSI, Xu2024NFeast} on the linear stability of the 1/1 MHD instabilities and the amplitude of the nonlinear dynamo term, as well as the spatial and velocity-space redistributions of energetic particles by the MHD instabilities and the dynamo electric field.}

\revisioncolor{The linear stabilities (growth rate and eigenfunction) of the n = 1 MHD mode for the axisymmetric equilibria at the quasi-stationary stage of the 3D JOREK simulations will be further analyzed at different parameters using CASTOR3D (a linear extended visco-resistive MHD code) \cite{Huysmans1993POFBcastor, Strumberger2017NFcastor3d}.  When combined with the fine-tuning of the saturated pressure profile and $q_0$ in JOREK nonlinear simulations by parameter scans (current/heating source, viscosity, resistivity, heat conductivity, etc.), this research would allow for the quantitative analyses of different driving effects (pressure gradient and current density) in the dominant MHD mode and the nonlinear dynamo term.} All parameter scans and the extended MHD developments are important for understanding the flux pumping mechanism through direct 3D MHD simulations and are also crucial for calibrating a fast surrogate model being developed \cite{Krebs2023EPSreduced}, which aims to efficiently predict the amplitude of dynamo electric field and the feasibility of flux pumping in both existing tokamaks like AUG \cite{Burckhart2023NF} and JET \cite{Burckhart2024EPS}, and future larger devices like ITER and DEMO. 

The necessity of the surrogate model \cite{Krebs2023EPSreduced} arises mainly from the enormous computational demand of initial value simulations. The presented 3D MHD simulation of flux pumping for the AUG discharge involves a significant timescale separation between the resistive diffusion time ($\tau_R$) and the Alfvén time ($\tau_A$). Specifically, the scale separation value of $\tau_R/\tau_A$ is greater than $10^6$ for AUG, which approximately requires 0.5 million core-hours for a 3D case on Marconi-Fusion supercomputer. In larger devices, the timescale separation typically increases due to the larger spatial scale, higher electron temperature, and stronger magnetic field. Specifically, with the parameters of the candidate discharges for flux pumping in JET \cite{Burckhart2024EPS}, the timescale separation is estimated to be 30\% larger than the AUG case. Thus, the 3D MHD simulation for flux pumping in JET is still feasible and is one of the main objectives of future JOREK simulations. However, for typical ITER and DEMO plasmas \cite{Sips2005PPCFITER, Siccinio2022DEMO}, the timescale separations involved are more than 10 and 40 times larger than the AUG situation, respectively. Simulations for larger tokamaks also require higher grid resolution. Therefore, 3D nonlinear MHD simulations of flux pumping for ITER and DEMO are impractical based on current computational resources, and the surrogate model \cite{Krebs2023EPSreduced}, after careful calibration, is essential for further assessing flux pumping in these reactor-scale devices.

In this context, the present simulation work represents an important milestone in successfully addressing the full-cycle nonlinear MHD behaviour of flux pumping in the experimental hybrid scenario with fully realistic plasma parameters, providing for the first time a quantitative agreement with experimental observations of flux pumping. Further studies regarding parameter scans and extended MHD developments are being carried out and will be reported in future publications.

\section{Acknowledgments}
The author Haowei Zhang would like to acknowledge Qingquan Yu, Daniele Bonfiglio, and Jörg Stober for valuable discussions, and Stanislas Pamela, Felix Antlitz, and Weikang Tang for assistance in implementing JOREK for this work. 
The simulations were finished mainly on the Marconi-Fusion supercomputer hosted by CINECA in Italy and the JFRS-1 supercomputer at IFERC-CSC in Japan.

This work has been carried out within the framework of the EUROfusion Consortium, funded by the European Union via the Euratom Research and Training Programme (Grant Agreement No 101052200 — EUROfusion). Views and opinions expressed are however those of the author(s) only and do not necessarily reflect those of the European Union or the European Commission. Neither the European Union nor the European Commission can be held responsible for them.

\bibliography{ref.bib}

\end{document}